\documentclass[aps,prd,10pt,nofootinbib,OliveGeqsecnum,showpacs,showkeys,superscriptaddress,preprintnumbers,altaffilletter,floatfix,twocolumn]{revtex4-2}

\usepackage{graphicx,caption,subcaption,bm,color,hyperref}
\usepackage{amsmath,amssymb}
\usepackage{mathtools} 
\usepackage{dsfont} 
\usepackage{multirow} 
\usepackage{float} 
\usepackage{mathrsfs} 
\usepackage{tensor} 
\usepackage{booktabs}
\usepackage{ragged2e}
\usepackage{needspace}   
\usepackage{placeins}    
\usepackage{afterpage}

\hypersetup{colorlinks, linkcolor={blue}, citecolor={red}, urlcolor={teal}}

\definecolor{amaranth}{rgb}{0.9, 0.17, 0.31}
\definecolor{palatinateblue}{rgb}{0.15, 0.23, 0.89}
\definecolor{brightpink}{rgb}{0.86, 0.26, 0.55}

\definecolor{forestgreen}{rgb}{0.13, 0.55, 0.13}
\definecolor{goldenrod}{rgb}{0.85, 0.65, 0.13}
\definecolor{darkviolet}{rgb}{0.58, 0.0, 0.83}
\definecolor{turquoise}{rgb}{0.0, 0.78, 0.8}

\hypersetup{ 
    linktoc=all,
    colorlinks, 
    linkcolor={palatinateblue},
    citecolor={brightpink}, 
    urlcolor={amaranth}
}

\graphicspath{{images/}}

\newcommand{\be}{\begin{equation}}
\newcommand{\ee}{\end{equation}}
\newcommand{\ba}{\begin{eqnarray}}
\newcommand{\ea}{\end{eqnarray}}

\begin{document}

\title{Modified Teleparallel $f(T)$ Gravity, DESI BAO  and the $H_0$ Tension}

\author{Mariam Bouhmadi-López}
\email{mariam.bouhmadi[at]ehu.eus}
\affiliation{IKERBASQUE, Basque Foundation for Science, 48011, Bilbao, Spain}
\affiliation{Department of Physics, University of the Basque Country UPV/EHU, P.O. Box 644, 48080 Bilbao, Spain}
\affiliation{EHU Quantum Center, University of the Basque Country UPV/EHU, P.O. Box 644, 48080 Bilbao, Spain}

\author{Carlos G. Boiza}
\email{carlos.garciab[at]ehu.eus}
\affiliation{Department of Physics, University of the Basque Country UPV/EHU, P.O. Box 644, 48080 Bilbao, Spain}
\affiliation{EHU Quantum Center, University of the Basque Country UPV/EHU, P.O. Box 644, 48080 Bilbao, Spain}

\author{Maria Petronikolou}
\email{petronikoloum[at]gmail.com}
\affiliation{Department of Physics, National Technical University of Athens, 
Zografou
Campus GR 157 73, Athens, Greece}
\affiliation{Institute for Astronomy, Astrophysics, Space Applications and 
Remote Sensing, National Observatory of Athens, 15236 Penteli, Greece}

\author{Emmanuel N. Saridakis}
\email{msaridak[at]noa.gr}
\affiliation{Institute for Astronomy, Astrophysics, Space Applications and 
Remote Sensing, National Observatory of Athens, 15236 Penteli, Greece}
\affiliation{Departamento de Matem\'{a}ticas, Universidad Cat\'{o}lica del 
Norte, Avda. Angamos 0610, Casilla 1280 Antofagasta, Chile}

\affiliation{CAS Key Laboratory for Researches in Galaxies and Cosmology, 
Department of Astronomy, University of Science and Technology of China, Hefei, 
Anhui 230026, P.R. China}

\begin{abstract}
We investigate whether late-time modifications of gravity in the teleparallel framework
can impact the current tension in the Hubble constant $H_0$, focusing on $f(T)$ cosmology
as a minimal and well-controlled extension of General Relativity. We consider three
representative $f(T)$ parametrisations that recover the teleparallel equivalent of General
Relativity at early times and deviate from it only at late epochs. 
The models are confronted with unanchored Pantheon+ Type~Ia supernovae, DESI DR2 baryon
acoustic oscillations, compressed Planck cosmic microwave background distance priors, and
redshift-space distortion data, allowing us to jointly probe the background expansion and
the growth of cosmic structures. Two of the three models partially shift the inferred value
of $H_0$ towards local measurements, while the third worsens the discrepancy. This behaviour
is directly linked to the effective torsional dynamics, with phantom-like regimes favouring
higher $H_0$ and quintessence-like regimes producing the opposite effect.
A global statistical comparison shows that the minimal $f(T)$ extensions considered here
are not favoured over $\Lambda$CDM by the combined data. Nevertheless, our results
demonstrate that late-time torsional modifications can non-trivially redistribute current
cosmological tensions among the background and growth sectors.
\end{abstract}

\maketitle

\renewcommand{\tocname}{Index}


\section{Introduction}\label{intro}

A wide range of cosmological observations indicate that the present Universe is undergoing an accelerated phase of expansion, characterised by the dominance of an effective component with negative pressure in the late-time energy budget. This conclusion is primarily supported by luminosity distance measurements of Type~Ia supernovae \cite{SupernovaSearchTeam:1998fmf,SupernovaCosmologyProject:1998vns,Bahcall:1999xn}, which provide direct evidence for a transition from decelerated to accelerated expansion, and is further reinforced by observations of the cosmic microwave background through distance priors and the angular scale of the acoustic peaks \cite{Bennett2003,Komatsu2011,Planck2013,Planck:2018vyg}, as well as by baryon acoustic oscillation measurements in the large-scale distribution of galaxies \cite{Eisenstein2005,Cole2005,Percival2010,Tegmark2004,Reid2010}. Additional support is provided by independent determinations of the expansion history from cosmic chronometers \cite{Jimenez:2001gg,Moresco:2024wmr} and local measurements of the Hubble parameter \cite{Riess:2020fzl}, which together place stringent constraints on the late-time dynamics of the Universe.

In order to account for this observationally established accelerated expansion, the standard cosmological framework models the large-scale dynamics of the Universe within the Friedmann–Lemaître–Robertson–Walker (FLRW) spacetime by introducing a dark energy component with an equation of state parameter close to $w=-1$. In the concordance $\Lambda$CDM paradigm \cite{Peebles:2002gy}, this role is played by a cosmological constant, which successfully reproduces the observed expansion history across a wide range of redshifts. Despite its phenomenological success, however, the cosmological constant suffers from well-known theoretical difficulties, including the fine-tuning and coincidence problems \cite{Weinberg:1988cp,Copeland:2006wr}. Moreover, the increasing precision of cosmological observations has revealed persistent tensions within the standard model, most notably the discrepancy between early- and late-Universe determinations of the Hubble constant \cite{CosmoVerse:2025txj}. These challenges have motivated the exploration of alternative scenarios, such as dynamical dark energy models \cite{Amendola:2015ksp} and modified theories of gravity \cite{CANTATA:2021asi}, in which additional degrees of freedom or deviations from General Relativity can give rise to late-time cosmic acceleration.

Modified theories of gravity provide a compelling and well-motivated framework to describe the late-time accelerated expansion of the Universe without resorting to a cosmological constant or an exotic dark energy component \cite{CANTATA:2021asi}. In these scenarios, cosmic acceleration emerges from deviations from General Relativity at large scales or low curvatures, leading to modified gravitational dynamics that can effectively generate a negative pressure component at late times \cite{Carroll:2000fy,Nojiri:2003ft}. Such theories allow for departures from the standard expansion history while remaining compatible with local gravity tests through suitable screening mechanisms \cite{Clifton:2011jh,DeFelice:2010aj}. Moreover, they offer new perspectives on outstanding theoretical and observational challenges, including the fine-tuning and coincidence problems associated with the cosmological constant, as well as current tensions between early- and late-Universe determinations of cosmological parameters \cite{CosmoVerse:2025txj}. As a result, modified gravity constitutes a versatile approach for extending the standard cosmological model and interpreting late-time observations in a unified manner \cite{CANTATA:2021asi}.

From a theoretical standpoint, such extensions can be systematically constructed by generalising the gravitational action beyond its linear dependence on the Ricci scalar. In curvature-based formulations, this leads to $f(R)$ gravity \cite{Capozziello:2011et,Nojiri:2010wj,Capozziello:2007ec,Zhang:2026sxi}, where the action is promoted to an arbitrary function of the Ricci scalar, giving rise to modified field equations and rich cosmological dynamics \cite{Morais:2015ooa}. Equivalent but conceptually distinct formulations can also be developed within teleparallel and symmetric teleparallel geometries, where gravitation is described by torsion or non-metricity, respectively, instead of curvature. In this context, $f(T)$ gravity generalises the teleparallel equivalent of General Relativity by replacing the torsion scalar with an arbitrary function \cite{Cai:2015emx,Bahamonde:2021gfp}, while $f(Q)$ gravity extends the symmetric teleparallel formulation through a general function of the non-metricity scalar \cite{BeltranJimenez:2017tkd,BeltranJimenez:2019tme}. In recent years, their cosmological applications have been explored in detail, both for $f(T)$ \cite{Nesseris:2013jea,Izumi:2012qj,Nunes:2016qyp,Golovnev:2020las,Mirza:2017vrk,Deng:2018ncg,Darabi:2019qpz,Zhao:2024uzq,Aquino:2025sdb, Tzerefos:2023mpe,Zhang:2021kqn} and for $f(Q)$ gravity \cite{Anagnostopoulos:2021ydo,Lazkoz:2019sjl,Lu:2019hra,Mandal:2020buf,Ayuso:2020dcu,Frusciante:2021sio,Gadbail:2022jco,Barros:2020bgg,Shabani:2023xfn,De:2023xua,Dimakis:2021gby,Anagnostopoulos:2022gej,Guzman:2024cwa,Heisenberg:2023lru,Boiza:2025xpn,Ayuso:2025vkc,Ferreira:2023awf,Su:2024avk}.

Even broader classes of models can be constructed by incorporating boundary terms that establish explicit connections between torsion- and non-metricity-based formulations of gravity, leading to generalised theories such as $F(Q,\mathcal{B})$ \cite{Capozziello:2023vne,De:2023xua} and $f(T,\mathcal{B})$ \cite{Bahamonde:2015zma,Bahamonde:2019shr}. These constructions provide a unified geometric description within the teleparallel and symmetric teleparallel frameworks, enlarging the space of viable cosmological models and offering a flexible setting for phenomenological studies of late-time cosmic acceleration.

In this work, we focus on $f(T)$ gravity and investigate three distinct models aimed at describing the late-time accelerated expansion of the Universe. We perform a comprehensive observational analysis by confronting their predictions with current cosmological data sets, including Type~Ia supernovae, baryon acoustic oscillations, the cosmic microwave background, and redshift-space distortions, allowing us to jointly probe the background expansion and the growth of cosmic structures. The models are constrained using a Bayesian Markov Chain Monte Carlo analysis, and their statistical performance is assessed relative to $\Lambda$CDM through information criteria.

The paper is organised as follows. Section~\ref{fTtheory} reviews $f(T)$ gravity and its formulation in a cosmological context within a FLRW Universe containing radiation, baryonic and dark matter. In Section~\ref{fTmodel}, we introduce the specific $f(T)$ models analysed in this work. Section~\ref{cosmo_data} presents the cosmological data sets employed and the statistical methodology used to constrain the model parameters. Finally, the main results and conclusions are summarised in Section~\ref{conclusions}.

\section{\texorpdfstring{$f(T)$} {f(T)} gravity and cosmology}
\label{fTtheory}

This section is devoted to a brief review of $f(T)$ gravity and its implementation within a cosmological framework.

\subsection{\texorpdfstring{$f(T)$} {f(T)} modified gravity}

Teleparallel gravity provides an equivalent formulation of General Relativity in which gravitation is described by spacetime torsion rather than curvature. The fundamental dynamical variables are the tetrad fields $e^{A}{}_{\mu}$, which define the spacetime metric through
\begin{equation}
g_{\mu\nu} = \eta_{AB}\, e^{A}{}_{\mu}\, e^{B}{}_{\nu},
\end{equation}
where $\eta_{AB}=\mathrm{diag}(-1,1,1,1)$ is the Minkowski metric. In this framework, gravitational interactions are encoded in the torsion tensor constructed from the Weitzenboeck connection\footnote{It is possible to formulate teleparallel gravity using a general affine connection defined in terms of both the tetrad field and a spin connection \cite{Golovnev:2017dox}. This covariant approach avoids the issue of local Lorentz violation that can arise in the standard $f(T)$ formulation. However, since the background and linear perturbation equations relevant for our analysis coincide in both approaches, we adopt the standard formulation for simplicity.} 
\begin{equation}
T^{\rho}{}_{\mu\nu} = e_{A}{}^{\rho}
\left(
\partial_{\mu} e^{A}{}_{\nu}
- \partial_{\nu} e^{A}{}_{\mu}
\right),
\end{equation}
which is curvature free but exhibits non vanishing torsion.

From the torsion tensor one defines the torsion scalar
\begin{equation}
T = S_{\rho}{}^{\mu\nu}\, T^{\rho}{}_{\mu\nu},
\end{equation}
where $S_{\rho}{}^{\mu\nu}$ is the superpotential constructed from the torsion tensor and its contractions. The teleparallel equivalent of General Relativity (TEGR) is obtained from an action linear in $T$ and leads to field equations that are dynamically equivalent to those of General Relativity. In particular, variation of the TEGR action with respect to the tetrad fields yields equations that can be written in the standard Einstein form,
\begin{equation}
G_{\mu\nu} = 8\pi G \, T_{\mu\nu},
\end{equation}
where $G_{\mu\nu}$ denotes the Einstein tensor associated with the Levi--Civita connection and $T_{\mu\nu}$ is the energy--momentum tensor of matter.

As in General Relativity, diffeomorphism invariance and minimal coupling imply the covariant conservation of the energy--momentum tensor,
\begin{equation}
\nabla_{\mu} T^{\mu\nu} = 0,
\end{equation}
ensuring local conservation of energy and momentum.

Extensions of this framework are obtained by promoting the torsion scalar in the gravitational action to an arbitrary function, giving rise to $f(T)$ gravity. The action of $f(T)$ gravity minimally coupled to matter is given by \cite{Cai:2015emx}
\begin{equation}
S = \frac{1}{16\pi G}
\int d^{4}x \, e \, f(T)
+ \int d^{4}x \, e \, \mathcal{L}_{\mathrm{m}},
\end{equation}
where $e=\det(e^{A}{}_{\mu})=\sqrt{-g}$ and $\mathcal{L}_{\mathrm{m}}$ denotes the matter Lagrangian.

Varying this action with respect to the tetrad fields yields the $f(T)$ field equations,

\begin{align}
&e^{-1}\partial_{\mu} \!\left( e\, e_{A}{}^{\rho} S_{\rho}{}^{\mu\nu} \right) f_{T}
+ e_{A}{}^{\rho} S_{\rho}{}^{\mu\nu}\, (\partial_{\mu}T)\, f_{TT} \nonumber\\[0.5em]
&\qquad - f_{T}\, e_{A}{}^{\lambda} T^{\rho}{}_{\mu\lambda} S_{\rho}{}^{\nu\mu}+ \frac{1}{4} e_{A}{}^{\nu} f=4\pi G\, e_{A}{}^{\rho} T_{\rho}{}^{\nu},
\end{align}
where $f_{T}\equiv df/dT$ and $f_{TT}\equiv d^{2}f/dT^{2}$. Despite the non linear dependence on the torsion scalar, the resulting field equations remain of second order in derivatives of the tetrad fields. The matter sector continues to satisfy the standard conservation equation,
\begin{equation}
\nabla_{\mu} T^{\mu\nu} = 0,
\end{equation}
as a consequence of diffeomorphism invariance and minimal coupling.

\subsection{\texorpdfstring{$f(T)$} {f(T)} cosmology}

We next introduce the cosmological framework of $f(T)$ gravity, first at the level of the background evolution and then within a linear perturbation analysis, which allows for a direct comparison with cosmological observations.

\subsubsection{\texorpdfstring{$f(T)$} {f(T)} cosmology in a FLRW universe}

In a cosmological context, we consider a spatially flat FLRW spacetime described by the line element
\begin{equation}
ds^{2} = -dt^{2} + a^{2}(t)\left(dx^{2}+dy^{2}+dz^{2}\right),
\end{equation}
where $a(t)$ is the scale factor and $H=\dot{a}/a$ denotes the Hubble parameter. For this background geometry, the torsion scalar reduces to
\begin{equation}
T = -6H^{2}.
\end{equation}

Assuming a Universe filled with pressureless baryonic matter, cold dark matter, and radiation, the cosmological field equations of $f(T)$ gravity lead to modified Friedmann equation and Raychaudhuri equation governing the background expansion. These equations can be written as \cite{Bengochea:2008gz,Wu:2010xk}
\begin{equation}
\label{feq1}
H^{2} = \frac{8\pi G}{3} \left(\rho_{\mathrm{b}} + \rho_{\mathrm{cdm}} + \rho_{\mathrm{r}} \right)
+ \frac{1}{6} \left(2T f_{T} - f - T\right),
\end{equation}
\begin{equation}
\label{feq2}
\dot{H} = -\frac{4\pi G} {\,2T f_{TT} + f_{T}}
\left( \rho_{\mathrm{b}} + \rho_{\mathrm{cdm}} + \rho_{\mathrm{r}} + p_{\mathrm{r}}\right),
\end{equation}
where $\rho_{\mathrm{r}}$ and $p_{\mathrm{r}} = \rho_{\mathrm{r}}/3$ denote the energy density and pressure of radiation, respectively; $\rho_{\mathrm{b}}$ and $\rho_{\mathrm{cdm}}$ stand for the energy densities of baryonic matter and cold dark matter; and $f_{T} \equiv df/dT$ and $f_{TT} \equiv d^{2}f/dT^{2}$. These expressions make explicit how deviations from the teleparallel equivalent of General Relativity modify the cosmic expansion through the torsional sector.

It is often convenient to recast the modified Friedmann equations into a form analogous to the standard cosmological equations by introducing an effective fluid description for the torsional contributions. In this representation, the Friedmann equations take the familiar form
\begin{equation}
H^{2} = \frac{8\pi G}{3}
\left(
\rho_{\mathrm{b}} + \rho_{\mathrm{cdm}} + \rho_{\mathrm{r}} + \rho_{T}
\right),
\end{equation}
\begin{equation}
\dot{H} =
-4\pi G
\left(
\rho_{\mathrm{b}} + \rho_{\mathrm{cdm}} + \rho_{\mathrm{r}}
+ p_{\mathrm{r}} + \rho_{T} + p_{T}
\right),
\end{equation}
where $\rho_{T}$ and $p_{T}$ represent the effective energy density and pressure associated with the torsional modifications introduced by $f(T)$ gravity.

The effective torsional energy density and pressure are defined as \cite{Linder:2010py,Bamba:2010wb}
\begin{equation}\label{rho}
\rho_{T} =\frac{1}{16\pi G}\left(2T f_{T} - f - T\right),
\end{equation}

\begin{equation}\label{pressuref(T)}
p_{T} = -\frac{1}{16\pi G} \left[-4\dot{H}\left( 2T f_{TT} + f_{T} - 1 \right)+ 2T f_{T} - f - T\right],
\end{equation}
which allow the torsional sector to be interpreted as an effective dark energy component.

It is then convenient to characterise the dynamical properties of this effective fluid through an equation of state parameter,
\begin{equation}\label{w_T}
w_{T} \equiv \frac{p_{T}}{\rho_{T}},
\end{equation}
which is, in general, time dependent and determined by the specific functional form of $f(T)$. For suitable choices of $f(T)$, the effective equation of state can drive late time accelerated expansion, providing a geometrical explanation of cosmic acceleration without the need for an explicit dark energy component.

For the observational analysis presented in the next section, it is convenient to introduce the dimensionless Hubble rate,
\begin{equation}
E(z)\equiv \frac{H(z)}{H_0},
\end{equation}
and the present-day density parameters
\begin{equation}
\Omega_{i0}\equiv \frac{8\pi G}{3H_0^2}\rho_{i0},
\end{equation}
where the subscript $i0$ labels the present-day values of the energy densities of the various cosmological components, the modified Friedmann equation in $f(T)$ gravity can be written as
\begin{equation}
E^{2}(z)=\Omega_{\mathrm{b}0}(1+z)^{3} +\Omega_{\mathrm{cdm}0}(1+z)^{3} +\Omega_{\mathrm{r}0}(1+z)^{4}+\Omega_{T}(z),
\end{equation}
where the effective torsional density parameter is defined as
\begin{equation}
\Omega_{T}(z)\equiv\frac{1}{6H_0^2}\left(2T f_{T}-f-T\right), \qquad T=-6H_0^2 E^2(z).
\end{equation}

\subsubsection{Linear cosmological perturbations in \texorpdfstring{$f(T)$} {f(T)} cosmology}

We now briefly discuss scalar cosmological perturbations in $f(T)$ gravity \cite{Chen:2010va,Izumi:2012qj,Golovnev:2018wbh} and derive the linear growth equation for matter density fluctuations \cite{Zheng:2010am,Souza:2024qwd}. This allows for a direct comparison between the perturbative behaviour of torsional and non-metricity-based modified gravity theories.

We consider scalar perturbations around a spatially flat FLRW background in the Newtonian gauge,
\begin{equation}
ds^{2}=-(1+2\Psi)\,dt^{2}
+a^{2}(t)(1-2\Phi)\,\delta_{ij}\,dx^{i}dx^{j},
\end{equation}
where $\Psi$ and $\Phi$ are the gauge-invariant Bardeen potentials. In general modified gravity theories, these two potentials need not coincide, leading to the presence of gravitational slip.

On subhorizon scales, $aH \ll k $, and within the quasi-static approximation, the linearised field equations of $f(T)$ gravity lead to modified Poisson equations of the form
\begin{equation}
k^{2}\Psi = -4\pi G_{\rm eff}(a)\,a^{2}\,\rho_{\rm m}\,\delta_{\rm m},
\end{equation}
\begin{equation}
k^{2}\Phi = -4\pi G_{\rm eff}(a)\,\eta(a)\,a^{2}\,\rho_{\rm m}\,\delta_{\rm m},
\end{equation}
where $\rho_{\rm m}$ and $\delta_{\rm m}$ denote the background matter density and its density contrast, respectively. The quantity $G_{\rm eff}$ is the effective gravitational coupling, while
\begin{equation}
\eta(a)\equiv\frac{\Phi}{\Psi}
\end{equation}
is the gravitational slip parameter.

For $f(T)$ gravity, the effective gravitational coupling and slip parameter are given by\footnote{We are assuming a vanishing scalar
anisotropic stress at linear order \cite{Chen:2010va,Wu:2010xk}.}
\begin{equation}
\label{geff}
G_{\rm eff}(a)=\frac{G}{f_{T}},
\qquad
\eta(a)=1,
\end{equation}
where $f_{T}\equiv df/dT$ is evaluated on the homogeneous background. Therefore, although gravity is effectively modified relative to General Relativity, the absence of additional anisotropic stress at linear order implies that the two metric potentials remain equal, provided one works on subhorizon scales within the quasi-static approximation.

Combining the modified Poisson equation with the conservation equations for pressureless matter,
\begin{equation}
\dot{\delta}_{\rm m}+\frac{\theta_{\rm m}}{a}=0,
\qquad
\dot{\theta}_{\rm m}+H\theta_{\rm m}-\frac{k^{2}}{a}\Psi=0,
\end{equation}
where $\theta_{\rm m}$ is the divergence of the matter velocity field, one obtains the evolution equation for the matter density contrast \cite{Chen:2010va,Golovnev:2018wbh}
\begin{equation}
\ddot{\delta}_{\rm m}
+2H\dot{\delta}_{\rm m}
-4\pi G_{\rm eff}(a)\,\rho_{\rm m}\,\delta_{\rm m}=0 .
\end{equation}

Introducing derivatives with respect to $\ln a$, and defining primes as $d/d\ln a$, this equation can be written as
\begin{equation}
\delta_{\rm m}''+
\left(
2+\frac{H'}{H}
\right)\delta_{\rm m}'
-\frac{3}{2}\,
\Omega_{\rm m}(a)\,
\frac{G_{\rm eff}(a)}{G}\,
\delta_{\rm m}=0.
\end{equation}
This expression makes explicit that deviations from General Relativity enter the growth of cosmic structures exclusively through the effective gravitational coupling $G_{\rm eff}$, while the gravitational slip remains trivial.

\section{Specific $f(T)$ models and effective torsional fluid}
\label{fTmodel}

Having established the general framework of $f(T)$ gravity and its cosmological background equations, we now introduce the specific functional forms of $f(T)$ that will be analysed in this work.

Let us start defining the present value of the torsion scalar as
\begin{equation}
T_0 \equiv -6H_0^2 ,
\end{equation}
where $H_0$ is the Hubble constant today. By construction, all models reduce to the teleparallel equivalent of General Relativity (TEGR) in the early Universe and introduce deviations relevant only at late times.

We next present the models, we will be considering:

\begin{enumerate}
   
    \item \textbf{First model}: 
The first model is given by
\begin{equation}
\label{f1}
f_1(T)=T\,\mathrm{e}^{\lambda_1 T_0/T},
\end{equation}
with $\lambda_1$ a dimensionless parameter. This model was originally proposed in Ref.~\cite{Awad:2017yod} and subsequently analysed in detail in Ref.~\cite{Akarsu:2024nas}.
The corresponding torsional energy density can be written as
\begin{equation}
\rho_T^{(1)} =\frac{T}{16\pi G}\left[\mathrm{e}^{\lambda_1 T_0/T}\left(1-2\lambda_1\frac{T_0}{T}\right)-1\right],
\end{equation}
while the pressure follows directly from Eq.~(\ref{pressuref(T)}).

The effective gravitational coupling \eqref{geff} reads
\begin{equation}\label{g1}
\frac{G_{\rm eff}^{(1)}}{G}=\frac{1}{\mathrm{e}^{\lambda_1 T_0/T}\left(1-\lambda_1\frac{T_0}{T}\right)}.
\end{equation}
Imposing the modified Friedmann equation \eqref{feq1} at $z=0$ ($T=T_0$ and $E(0)=1$) fixes $\lambda_1$ through
\begin{equation}
\lambda_1=\frac{1}{2}+\mathcal{W}_0\!\left(-\frac{\Omega_{\mathrm{b0}}+\Omega_{\mathrm{cdm0}}+\Omega_{\mathrm{r0}}}{2\sqrt{e}}\right),
\end{equation}
where $\mathcal{W}_0$ is the Lambert function (principal branch for the cosmological solution).

\item \textbf{{Second model}}: The second model reads as \cite{Bamba:2010wb} 
\begin{equation}
\label{f2}
f_2(T)=T + T_0\,\mathrm{e}^{-\lambda_2 T_0/T},
\end{equation}
where $\lambda_2$ is a dimensionless constant. The effective torsional energy density becomes
\begin{equation}
\rho_T^{(2)} =\frac{T_0}{16\pi G}\,\mathrm{e}^{-\lambda_2 T_0/T}\left(1+2\lambda_2\frac{T_0}{T}\right),
\end{equation}
while the corresponding pressure is obtained directly from Eq.~(\ref{pressuref(T)}).

The effective gravitational coupling \eqref{geff} reads
\begin{equation}\label{g2}
\frac{G_{\rm eff}^{(2)}}{G}=\frac{1}{1+\lambda_2\left(\frac{T_0}{T}\right)^2\mathrm{e}^{-\lambda_2 T_0/T}}.
\end{equation}
Imposing the modified Friedmann equation \eqref{feq1} at $z=0$ yields
\begin{equation}
\lambda_2=\frac{1}{2}-\mathcal{W}_0\!\left[\frac{\sqrt{e}}{2}\left(1-\Omega_{\mathrm{b0}}-\Omega_{\mathrm{cdm0}}-\Omega_{\mathrm{r0}}\right)\right].
\end{equation}

\item \textbf{{Third model}}:
Finally, we consider a novel model, characterized by 
\begin{equation}
\label{f3}
f_3(T)=T+\lambda_3 T_0\left[1-\mathrm{e}^{-T_0/T}\right],
\end{equation}
with $\lambda_3$ a dimensionless parameter. The effective energy density of the torsional sector is
\begin{equation}
\rho_T^{(3)} =\frac{\lambda_3 T_0}{16\pi G}\left[1-\mathrm{e}^{-T_0/T}\left(1+2\frac{T_0}{T}\right)\right],
\end{equation}
while the associated pressure follows directly from Eq.~(\ref{pressuref(T)}).

The effective gravitational coupling \eqref{geff} reads
\begin{equation}\label{g3}
\frac{G_{\rm eff}^{(3)}}{G}=\frac{1}{1-\lambda_3\left(\frac{T_0}{T}\right)^2\mathrm{e}^{-T_0/T}}.
\end{equation}
At $z=0$ the Friedmann equation \eqref{feq1} gives the closed form
\begin{equation}
\lambda_3=\frac{e}{1+e}\left(1-\Omega_{\mathrm{b0}}-\Omega_{\mathrm{cdm0}}-\Omega_{\mathrm{r0}}\right).
\end{equation}
\end{enumerate}

For all three models, the background evolution is fully determined once the modified Friedmann equations are solved for $H(z)$. The redshift dependence of the effective equation of state $w_T(z)$ provides a convenient diagnostic to assess whether the torsional sector can drive late time cosmic acceleration. 

\begin{figure*}[t!]
    \centering
    
    \begin{subfigure}[b]{0.5\textwidth}
        \centering
        \includegraphics[width=\textwidth, height=0.6\textheight, keepaspectratio]{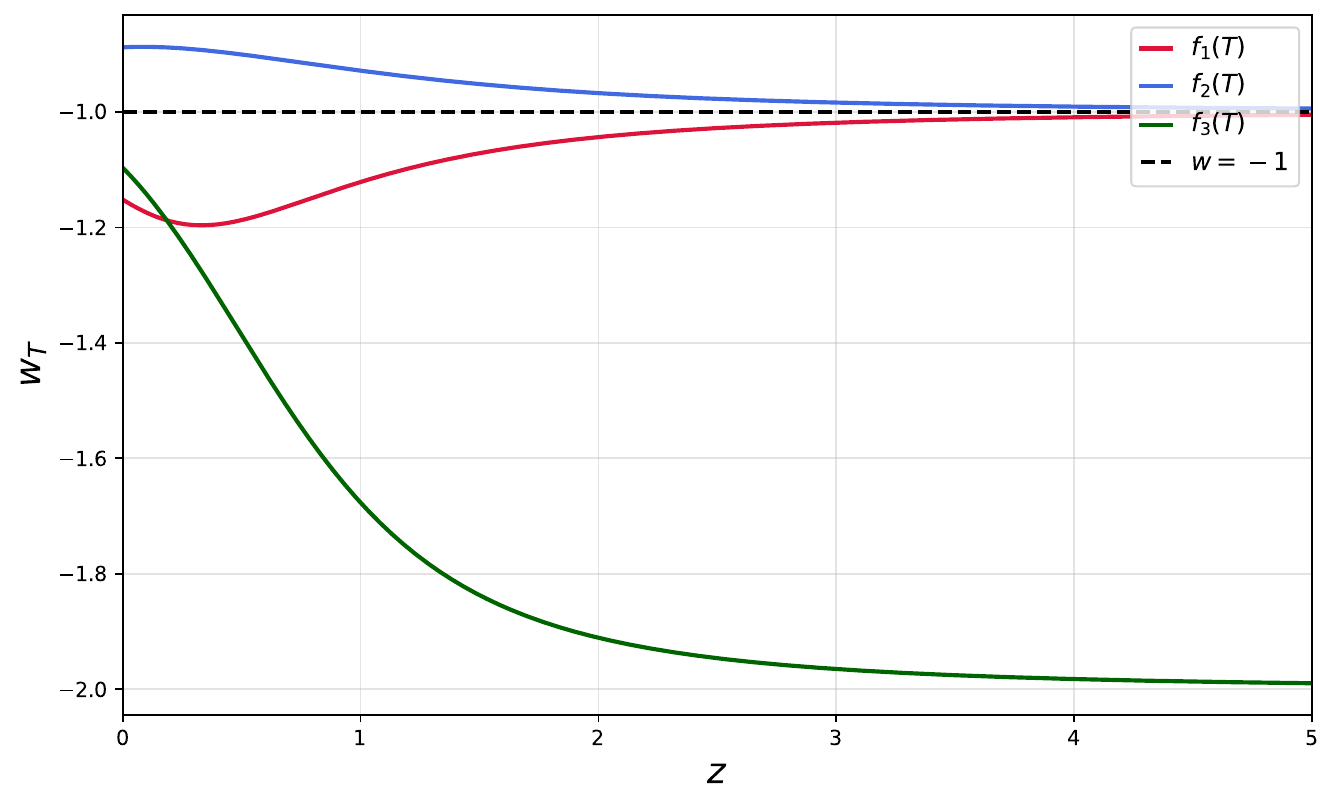}
        \label{fig:theory_wt}
    \end{subfigure}
    \hfill
    \begin{subfigure}[b]{0.49\textwidth}
        \centering
        \includegraphics[width=\textwidth, height=0.25\textheight, keepaspectratio]{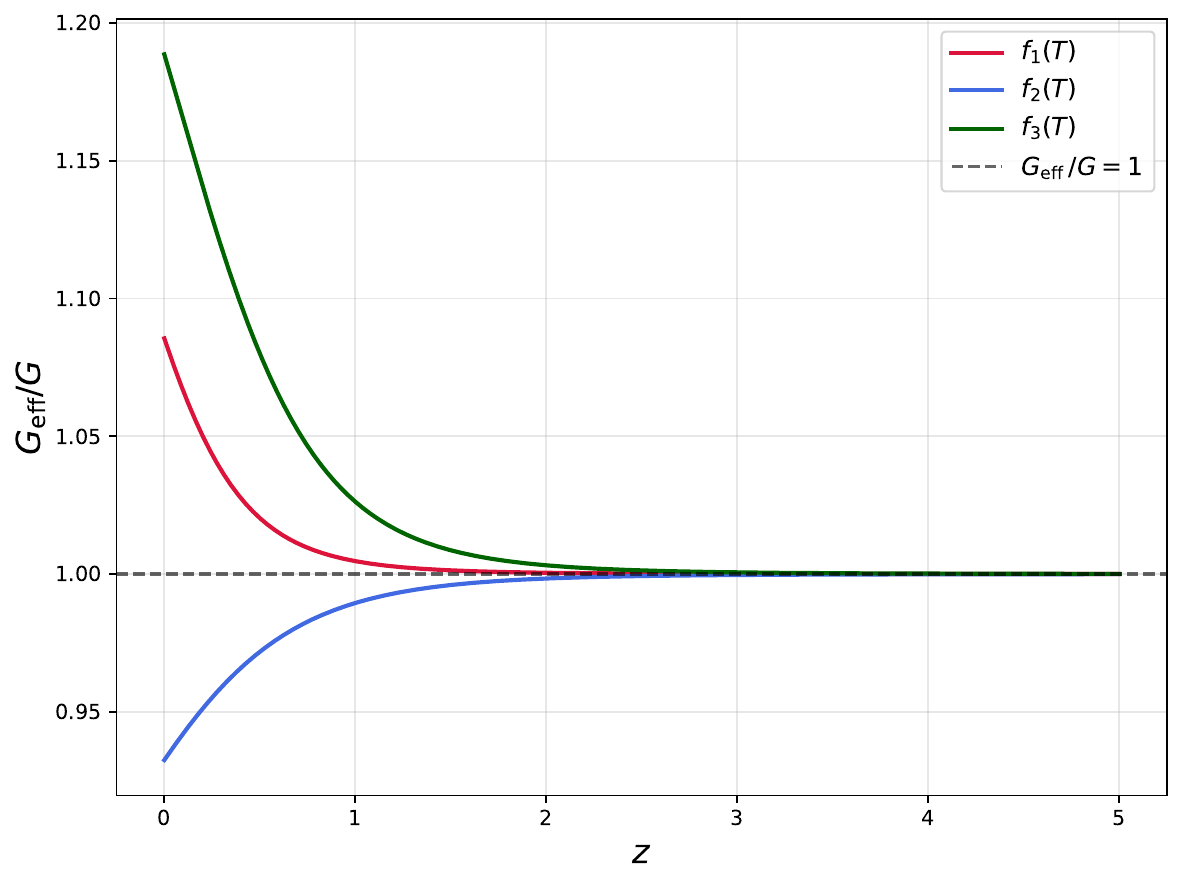}
        \label{fig:theory_g}
    \end{subfigure}

    \vspace{8pt} 

    \caption{\justifying{\textit{Cosmological background and effective sector of the $f(T)$ scenarios in comparison with $\Lambda$CDM. 
    \textbf{Left panel:} Effective dark energy equation of state parameter $w_T(z)$ of the $f(T)$ models compared to $\Lambda$CDM ($w=-1$). Models~1 (crimson line) and ~3 (green line) with $\lambda_1=0.36$ and $\lambda_3=0.42$ respectively, show phantom-like behaviour ($w_T<-1$), whereas Model~2 with $\lambda_2=0.09$ (blue line), lies in the quintessence regime ($w_T>-1$).
    \textbf{Right panel:} Effective Newton’s constant $G_{\mathrm{eff}}/G$ as a function of the redshift $z$. 
    Models~1 (crimson line) and ~3 (green line) with $\lambda_1=0.36,\, \lambda_3=0.42$ respectively, show $G_{\mathrm{eff}}>G$, whereas Model~2 with $\lambda_2=0.09$ (blue line) exhibits $G_{\mathrm{eff}}<G$, 
    with distinct implications for structure formation. The black dashed line $G_{\mathrm{eff}}/G=1$ 
    corresponds to the GR limit.}}}
    \label{fig:theory_wde}
\end{figure*}

In Fig.~\ref{fig:theory_wde}, we illustrate the redshift evolution of the effective
dark energy equation-of-state parameter $w_T(z)$ obtained from \eqref{w_T} using \eqref{pressuref(T)}, \eqref{rho} and the effective gravitational
coupling $G_{\rm eff}/G$ for the three $f(T)$ models, Eqs. \eqref{g1}, \eqref{g2} and \eqref{g3} introduced in this section.
For this purpose, we numerically solve the modified Friedmann equations derived
from the $f(T)$ field equations, assuming a spatially flat FLRW background
containing standard radiation and pressureless matter components.

The initial conditions are chosen such that all models recover the teleparallel
equivalent of General Relativity at early times, ensuring consistency with the
standard cosmological evolution during radiation and matter domination. As a
result, deviations from $\Lambda$CDM arise only at late times, when torsional
effects become dynamically relevant. The figure therefore serves as a qualitative
diagnostic of the intrinsic late-time behaviour associated with each
parametrisation, rather than as a fit to observational data.

As shown in the left panel of Fig.~\ref{fig:theory_wde}, Models~1 and~3 enter a phantom-like regime characterised by
$w_T<-1$ over the relevant redshift
range, whereas Model~2 exhibits an
effective quintessence-like behaviour, with $w_T>-1$. These distinct behaviours reflect the different functional forms of
$f(T)$ and determine the way in which torsional contributions modify the cosmic
expansion rate at late times. The corresponding impact on the gravitational
sector is displayed in the right panel, where Models~1 and~3 satisfy
$G_{\rm eff}>G$, indicating an enhancement of the effective gravitational
interaction, while Model~2 realises $G_{\rm eff}<G$, corresponding to a weakening
of gravity.

These qualitative differences play a central role in shaping both the background
expansion history and the growth of cosmic structures, and provide clear physical
intuition for the trends that will be identified in the subsequent observational
analysis. In particular, the sign and magnitude of the torsional contributions
anticipate whether a given model tends to shift cosmological parameter
constraints, such as the inferred value of the Hubble constant, towards or away
from their standard $\Lambda$CDM values. A quantitative analysis of these
effects will be presented in the following sections through a full comparison
with current cosmological data. 

\section{Cosmological Data and Parameter Estimation}
\label{cosmo_data}

In this section we confront the $f(T)$ models introduced above with current
cosmological observations in order to constrain their parameter spaces and
assess their phenomenological viability relative to the standard
$\Lambda$CDM scenario.  The analysis is designed to quantify both the degree to which the
models can accommodate late-time observational constraints and the impact of
torsional modifications on key cosmological parameters.

\subsection{Data and Methodology}
\label{datamethod}

We perform a Bayesian analysis to constrain the cosmological parameters of the
$f(T)$ models introduced in Section~\ref{fTmodel}. Posterior distributions are
obtained using the Monte Carlo Markov Chain (MCMC) sampler implemented in the
\texttt{Cobaya} framework \cite{Torrado:2020dgo}. For each data combination, we
carry out two independent MCMC runs using identical numerical settings and
convergence criteria. Convergence of each run is assessed using the
Gelman-Rubin $R-1$ statistic \cite{Lewis:2013hha}, after which the two converged
chains are combined to construct the final posterior distributions.

The statistical analysis is based on a Gaussian likelihood constructed from the
combined contribution of the cosmological probes considered in this work.
Depending on the data combination, this includes Type~Ia supernovae, baryon
acoustic oscillations, cosmic microwave background distance priors, and
redshift-space distortion measurements. The precise definition of each
likelihood component is described in the corresponding data subsection below.

\Needspace{8\baselineskip}

The set of sampled cosmological parameters is chosen to reflect the quantities
most directly constrained by the data. We sample the present-day Hubble constant
$H_0$, the physical cold dark matter density $\Omega_{\mathrm{cdm}0}h^2$, and the
physical baryon density $\Omega_{\mathrm{b}0}h^2$, where
$h\equiv H_0/(100~\mathrm{km\,s^{-1}\,Mpc^{-1}})$. When redshift-space distortion
data are included, we additionally sample the present-day amplitude of matter
fluctuations $\sigma_8$. The parameter vector is therefore given by
\begin{equation}
\bm{\theta} =
\begin{cases}
\{H_0,\,\Omega_{\mathrm{cdm}0}h^2,\,\Omega_{\mathrm{b}0}h^2\}, & \text{without RSD},\\[4pt]
\{H_0,\,\Omega_{\mathrm{cdm}0}h^2,\,\Omega_{\mathrm{b}0}h^2,\,\sigma_8\}, & \text{with RSD}.
\end{cases}
\end{equation}

We adopt conservative priors on all sampled parameters. Flat (uniform) priors
are assumed for the Hubble constant $H_0$, the physical cold dark matter density
$\Omega_{\mathrm{cdm}0}h^2$, and, when included, the amplitude of matter
fluctuations $\sigma_8$. For the physical baryon density
$\Omega_{\mathrm{b}0}h^2$, we adopt a Gaussian prior motivated by Big Bang
Nucleosynthesis (BBN) constraints. The adopted priors and their numerical values
are summarised in Table~\ref{tab:priors}.

\begin{table}[t]
\centering
\begin{tabular}{l c}
\hline\hline
Parameter & Prior \\
\hline
$H_0$ & $\mathcal{U}(20,\,100)$ \\
$\Omega_{\mathrm{cdm}0}h^2$ & $\mathcal{U}(0.001,\,0.99)$ \\
$\Omega_{\mathrm{b}0}h^2$ & $\mathcal{N}(0.0222,\,0.0005)$ \\
$\sigma_8$ & $\mathcal{U}(0,\,2)$ \\
\hline\hline
\end{tabular}
\caption{\justifying{Priors adopted for the sampled cosmological parameters.
Flat (uniform) priors are used for $H_0$, $\Omega_{\mathrm{cdm}0}h^2$, and
$\sigma_8$, while a Gaussian prior motivated by Big Bang Nucleosynthesis
constraints is imposed on $\Omega_{\mathrm{b}0}h^2$. The parameter $\sigma_8$ is
included only when redshift-space distortion data are used.}}
\label{tab:priors}
\end{table}

While the MCMC sampling is performed in terms of the above parameters, the
results are reported in terms of the present-day Hubble constant $H_0$, the baryon density parameter $\Omega_{\mathrm{b}0}$, the
total matter density parameter
\begin{equation}
\Omega_{\mathrm{m}0}\equiv\Omega_{\mathrm{cdm}0}+\Omega_{\mathrm{b}0},
\end{equation}
and the derived clustering amplitude
\begin{equation}
S_8 \equiv \sigma_8\sqrt{\frac{\Omega_{\mathrm{m}0}}{0.3}}.
\end{equation}
These quanities provide a more transparent physical interpretation of the
results and allow for a direct comparison with constraints from other
cosmological analyses.

The radiation density parameter is fixed by standard early-Universe physics and
is not treated as a free parameter. It is determined through the redshift of
matter--radiation equality \cite{Eisenstein:1997ik},
\begin{equation}
z_{\mathrm{eq}} =
2.5\times 10^4\,\Omega_{\mathrm{m}0}h^2
\left(\frac{T_{\mathrm{CMB}}}{2.7}\right)^{-4},
\quad
\Omega_{\mathrm{r}0}=\frac{\Omega_{\mathrm{m}0}}{1+z_{\mathrm{eq}}},
\end{equation}
with the CMB temperature fixed to $T_{\mathrm{CMB}}=2.7255~\mathrm{K}$
\cite{Fixsen:2009ug}.

\subsubsection{Type Ia Supernovae}
\label{sndata}

We use the Pantheon+ compilation of Type~Ia supernovae \cite{Brout:2022vxf},
consisting of 1701 spectroscopically confirmed events spanning the redshift
range $0.001 \lesssim z \lesssim 2.3$. Type~Ia supernovae act as standardisable
candles and provide precise measurements of the luminosity--distance relation,
thereby constraining the late-time expansion history of the Universe.

In this work, the Pantheon+ sample is treated as an \emph{unanchored} supernova
dataset. This means that supernovae constrain only the redshift dependence of
the luminosity distance and do not fix the absolute distance scale. Accordingly,
we analytically marginalise over the nuisance parameter $\mathcal{M}$,
corresponding to the absolute magnitude of Type~Ia supernovae, following the
prescription of \cite{SNLS:2011lii}. The full covariance matrix provided with
the Pantheon+ sample is used, accounting for both statistical and systematic
uncertainties.

In order to anchor the distance scale whenever supernova data are included in
the analysis, we impose a Gaussian prior on the Hubble constant based on local
distance-ladder measurements. We adopt the determination by Riess
\textit{et al.}~\cite{Riess:2020fzl},
\begin{equation}
H_0 = 73.2 \pm 1.3~\mathrm{km\,s^{-1}\,Mpc^{-1}}.
\end{equation}
This prior effectively calibrates the supernova absolute magnitude while
preserving the unanchored nature of the Pantheon+ dataset.

\subsubsection{Baryon Acoustic Oscillations}
\label{baodata}

Baryon acoustic oscillations (BAO) provide a robust geometrical probe of the
cosmic expansion history through a standard ruler set by the sound horizon.
BAO measurements constrain combinations of cosmological distance measures,
relative to the sound horizon scale at the drag epoch, $r_\mathrm{s}(z_{\mathrm{d}})$.

In this work, we employ the most recent BAO measurements from the DESI
Data Release~2 (DR2) \cite{DESI:2025zgx}, which cover a wide redshift range and
multiple tracers, including bright galaxies, luminous red galaxies, emission
line galaxies, quasars, and the Lyman-$\alpha$ forest. These data provide
high-precision constraints on the late-time expansion history and are
particularly powerful when combined with supernovae and CMB distance priors.

The sound horizon at the drag epoch, $r_\mathrm{s}(z_{\mathrm{d}})$, is computed using the
same fitting formula adopted in the DESI analyses \cite{DESI:2024mwx,DESI:2025zgx}, ensuring full consistency
with the treatment of early-Universe physics in the BAO likelihood. The full
covariance matrix provided by the DESI collaboration is used, accounting for
correlations among different redshift bins and tracers.

\subsubsection{Cosmic Microwave Background}
\label{cmbdata}

The Cosmic Microwave Background (CMB) provides a precise probe of the early
Universe and plays a key role in anchoring the cosmic distance scale. Rather
than using the full CMB temperature and polarisation power spectra, we adopt
compressed CMB distance priors derived from Planck observations
\cite{Zhai:2018vmm}.

These priors are expressed in terms of the shift parameters $R$ and $\ell_a$,
together with the physical baryon density $\Omega_{\mathrm{b}0}h^2$, and
efficiently capture the geometrical information relevant for late-time
cosmology. The CMB likelihood is constructed using the covariance matrix
provided in \cite{Zhai:2018vmm}. Since the $f(T)$ models considered here reduce
to standard cosmology at early times, the use of distance priors is well
justified.

\subsubsection{Redshift-Space Distortions}
\label{rsddata}

Redshift-space distortions (RSD) arise from the peculiar velocities of galaxies
and introduce anisotropies in the observed clustering pattern in redshift
space. RSD measurements constrain the growth of cosmic structures through the
quantity $f\sigma_8(z)$, defined as the product of the linear growth rate $f$
and the amplitude of matter fluctuations $\sigma_8$.

The growth rate is defined as
\begin{equation}\label{growthrate}
f(a)=\frac{d\ln\delta_{\mathrm{m}}}{d\ln a},
\end{equation}
where $\delta_{\mathrm{m}}$ is the matter density contrast, whose evolution is
obtained by numerically solving the linear growth equation derived in
Section~\ref{fTtheory}. The redshift evolution of $\sigma_8$ is given by \cite{Wang:2010gq}
\begin{equation}
\sigma_8(a)=\sigma_8\,\frac{\delta_{\mathrm{m}}(a)}{\delta_{\mathrm{m}}(a=1)},
\end{equation}
with $\sigma_8\equiv\sigma_8(a=1)$ treated as a free parameter.

We use a compilation of 22 $f\sigma_8(z)$ measurements spanning a wide redshift
range, as presented in \cite{Sagredo:2018ahx}. In this work, the RSD likelihood
is constructed assuming uncorrelated errors and is combined consistently with
the background probes described above.

\subsection{Information Criteria}
\label{infocrit}

To assess the relative performance of the $f(T)$ models with respect to the
standard $\Lambda$CDM scenario, we employ information-theoretic model selection
criteria that balance goodness of fit against model complexity.

In particular, we use the corrected Akaike Information Criterion
$\mathrm{AIC}_{\mathrm{C}}$ \cite{Akaike:1974vps,Kenneth:2004a,Kenneth:2004b},
defined as
\begin{equation}
\mathrm{AIC}_{\mathrm{C}} =
-2\ln L_{\mathrm{max}} + 2\kappa +
\frac{2\kappa(\kappa+1)}{N-\kappa-1},
\end{equation}
where $L_{\mathrm{max}}$ is the maximum likelihood achieved by the model, $\kappa$
denotes the number of free parameters, and $N$ is the total number of data
points. In the limit of large $N$, the correction term becomes negligible and
$\mathrm{AIC}_{\mathrm{C}}$ reduces to the standard Akaike Information Criterion,
$\mathrm{AIC}$ \cite{Liddle:2007fy}.

Since all models considered in this work (including $\Lambda$CDM and the $f(T)$
models) are constrained using the same datasets and have the same number of free
parameters, both $\kappa$ and $N$ are identical across models. As a result, the
difference in $\mathrm{AIC}_{\mathrm{C}}$ simplifies to
\begin{equation}
\begin{aligned}
\Delta\mathrm{AIC}_{\mathrm{C}}
&\equiv
\mathrm{AIC}_{\mathrm{C}}^{f(T)}
-
\mathrm{AIC}_{\mathrm{C}}^{\Lambda\mathrm{CDM}} \\
&=
-2\left[
\ln L^{f(T)}_{\mathrm{max}}
-
\ln L^{\Lambda\mathrm{CDM}}_{\mathrm{max}}
\right].
\end{aligned}
\end{equation}
Negative
values of $\Delta\mathrm{AIC}_{\mathrm{C}}$ indicate a preference for the $f(T)$
model, while positive values favour $\Lambda$CDM. The interpretation of
$\Delta\mathrm{AIC}_{\mathrm{C}}$ follows the Jeffreys’ scale \cite{Jeffreys:1961a}, summarised in
Table~\ref{tab:jeffreys}.

\begin{table}[t]
\centering
\begin{tabular}{c c}
\hline\hline
$\left|\Delta\mathrm{AIC}_{\mathrm{C}}\right|$ & Interpretation \\
\hline
$<2$        & Compatible \\
$2$--$5$    & Moderate evidence \\
$5$--$10$   & Strong evidence \\
$>10$       & Decisive evidence \\
\hline\hline
\end{tabular}
\caption{\justifying{Jeffreys’ scale for the interpretation of the absolute
difference $\left|\Delta\mathrm{AIC}_{\mathrm{C}}\right|$. The sign of $\Delta\mathrm{AIC}_{\mathrm{C}}$ determines
the preferred model: negative values favour the $f(T)$ model, while positive
values favour the reference $\Lambda$CDM scenario.}}
\label{tab:jeffreys}
\end{table}

For completeness, we note that the Bayesian Information Criterion (BIC)
\cite{Schwarz:1978tpv} leads to identical conclusions in this analysis, since all
models share the same number of parameters and are fitted to the same data.

\subsection{Results}
\label{rslts}

\begin{table*}[t]
\centering
\begin{tabular}{@{\hskip 11pt}c@{\hskip 11pt}c@{\hskip 11pt}c@{\hskip 11pt}c@{\hskip 11pt}c@{\hskip 11pt}c@{\hskip 11pt}c@{\hskip 11pt}}
\hline
\hline
\addlinespace[2pt]
Model 
& $H_0$ 
& $\Omega_{\mathrm{m}0}$ 
& $\Omega_{\mathrm{b}0}$ 
& $S_8$ 
& $\chi^2_{\min}$ 
& $\Delta\mathrm{AIC}_{\mathrm{C}}$ \\
\addlinespace[2pt]
\hline
\hline

\addlinespace[2pt]
\multicolumn{7}{c}{\textbf{SN}} \\
\addlinespace[2pt]

$\Lambda$CDM 
& $73.28 \pm 1.30$ 
& $0.3321 \pm 0.0186$ 
& $0.04138 \pm 0.00172$ 
& $-$ 
& $1402.9$
& $-$ \\

$f_1(T)$ 
& $73.29 \pm 1.26$ 
& $0.3870 \pm 0.0177$ 
& $0.04136 \pm 0.00175$ 
& $-$ 
& $1405.0$
& $2.1$ \\

$f_2(T)$ 
& $73.24 \pm 1.28$ 
& $0.2792 \pm 0.0136$ 
& $0.04144 \pm 0.00172$ 
& $-$  
& $1402.5$
& $-0.4$ \\

$f_3(T)$ 
& $73.24 \pm 1.28$ 
& $0.3862 \pm 0.0197$ 
& $0.04141 \pm 0.00174$ 
& $-$  
& $1403.5$
& $0.6$ \\

\addlinespace[2pt]
\hline

\addlinespace[2pt]
\multicolumn{7}{c}{\textbf{BAO}} \\
\addlinespace[2pt]

$\Lambda$CDM 
& $68.645 \pm 0.505$ 
& $0.29747 \pm 0.00861$ 
& $0.047173 \pm 0.000728$ 
& $-$ 
& $10.3$
& $-$ \\

$f_1(T)$ 
& $72.318 \pm 0.542$ 
& $0.29458 \pm 0.00785$ 
& $0.042475 \pm 0.000673$ 
& $-$ 
& $20.1$
& $9.8$ \\

$f_2(T)$ 
& $65.278 \pm 0.529$ 
& $0.30541 \pm 0.00956$ 
& $0.052180 \pm 0.000812$ 
& $-$ 
& $9.1$
& $-1.2$ \\

$f_3(T)$ 
& $72.654 \pm 0.553$ 
& $0.31003 \pm 0.00819$ 
& $0.042070 \pm 0.000645$ 
& $-$ 
& $20.4$
& $10.1$ \\

\addlinespace[2pt]
\hline

\addlinespace[2pt]
\multicolumn{7}{c}{\textbf{BAO + CMB}} \\
\addlinespace[2pt]

$\Lambda$CDM 
& $68.401 \pm 0.292$ 
& $0.30111 \pm 0.00374$ 
& $0.048133 \pm 0.000343$ 
& $-$ 
& $16.2$
& $-$ \\

$f_1(T)$ 
& $71.963 \pm 0.326$ 
& $0.27621 \pm 0.00350$ 
& $0.043151 \pm 0.000321$ 
& $-$ 
& $27.8$
& $11.6$ \\

$f_2(T)$ 
& $64.850 \pm 0.348$ 
& $0.33107 \pm 0.00483$ 
& $0.053884 \pm 0.000476$ 
& $-$ 
& $29.8$
& $13.6$ \\

$f_3(T)$ 
& $72.317 \pm 0.320$ 
& $0.27778 \pm 0.00345$ 
& $0.042406 \pm 0.000300$ 
& $-$ 
& $43.5$
& $27.3$ \\

\addlinespace[2pt]
\hline

\addlinespace[2pt]
\multicolumn{7}{c}{\textbf{RSD}} \\
\addlinespace[2pt]

$\Lambda$CDM 
& $-$ 
& $0.2695 \pm 0.0540$ 
& $-$
& $0.7423 \pm 0.0378$ 
& $11.94$
& $-$ \\

$f_1(T)$ 
& $-$ 
& $0.2604 \pm 0.0521$ 
& $-$
& $0.7104 \pm 0.0385$  
& $12.06$
& $0.12$ \\

$f_2(T)$ 
& $-$ 
& $0.2781 \pm 0.0559$ 
& $-$ 
& $0.7695 \pm 0.0430$ 
& $11.96$
& $0.02$ \\

$f_3(T)$ 
& $-$ 
& $0.2444 \pm 0.0499$ 
& $-$ 
& $0.6848 \pm 0.0371$  
& $12.02$
& $0.08$ \\

\addlinespace[2pt]
\hline

\addlinespace[2pt]
\multicolumn{7}{c}{\textbf{SN + BAO + CMB + RSD}} \\
\addlinespace[2pt]

$\Lambda$CDM 
& $68.559 \pm 0.278$ 
& $0.29923 \pm 0.00354$ 
& $0.048014 \pm 0.000326$ 
& $0.7582 \pm 0.0263$ 
& $1447.8$
& $-$ \\

$f_1(T)$ 
& $71.620 \pm 0.322$ 
& $0.28013 \pm 0.00352$ 
& $0.043475 \pm 0.000315$ 
& $0.7222 \pm 0.0253$ 
& $1486.8$
& $39.0$ \\

$f_2(T)$ 
& $65.699 \pm 0.316$ 
& $0.31955 \pm 0.00420$ 
& $0.052873 \pm 0.000426$ 
& $0.7966 \pm 0.0283$ 
& $1492.4$
& $44.6$ \\

$f_3(T)$ 
& $72.043 \pm 0.305$ 
& $0.28088 \pm 0.00335$ 
& $0.042649 \pm 0.000293$ 
& $0.7071 \pm 0.0249$ 
& $1494.1$
& $46.3$ \\

\addlinespace[2pt]
\hline
\hline
\end{tabular}
\caption{\justifying{
Mean values and $1\sigma$ uncertainties of the cosmological parameters for the $\Lambda$CDM and $f(T)$ gravity models, obtained from individual datasets (SN, BAO, RSD) and from the combined BAO+CMB dataset, as well as from their full combination.
For the RSD-only dataset, no constraints on $H_0$ and $\Omega_{\mathrm{b}0}$ are reported, because these parameters are not directly constrained by RSD measurements alone.
The column $\chi^2_{\min}$ reports the minimum chi-square value at the best-fit point.
The last column reports the Akaike Information Criterion difference,
$\Delta\mathrm{AIC}_{\mathrm{C}} \equiv \mathrm{AIC}_{\mathrm{C}}^{f(T)} - \mathrm{AIC}_{\mathrm{C}}^{\Lambda\mathrm{CDM}}$.
}}
\label{tab:fT_results}
\end{table*}

\begin{figure*}[t]  
    \centering
    \begin{tabular}{cc}
        \includegraphics[width=0.49\linewidth]{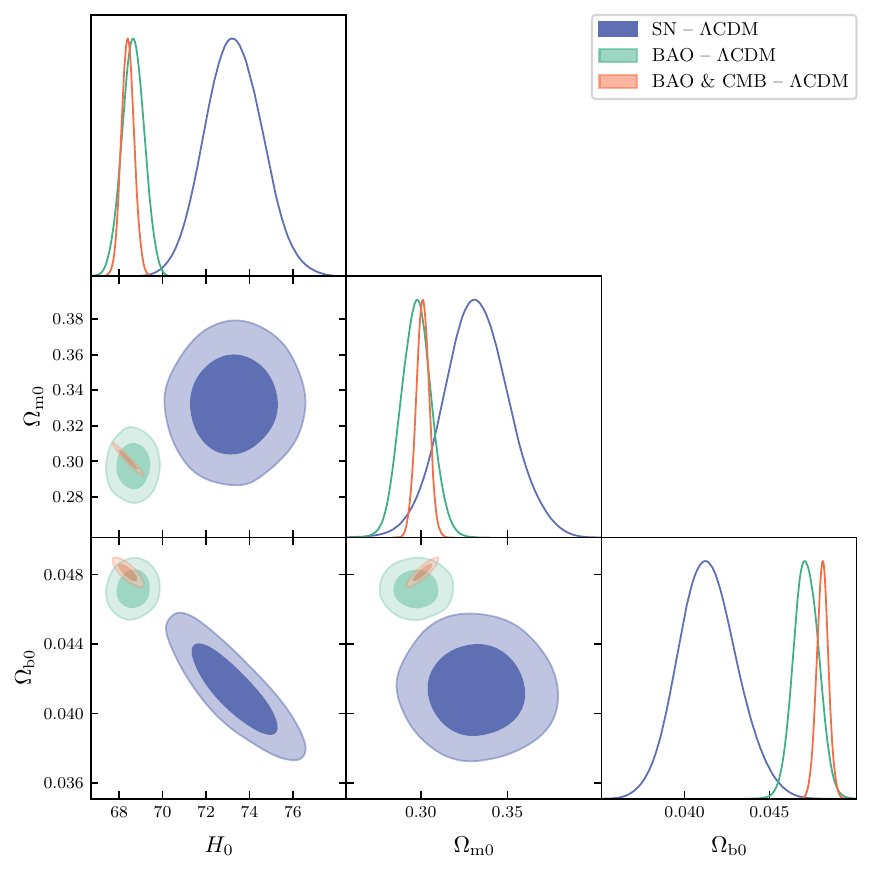} &
        \includegraphics[width=0.49\linewidth]{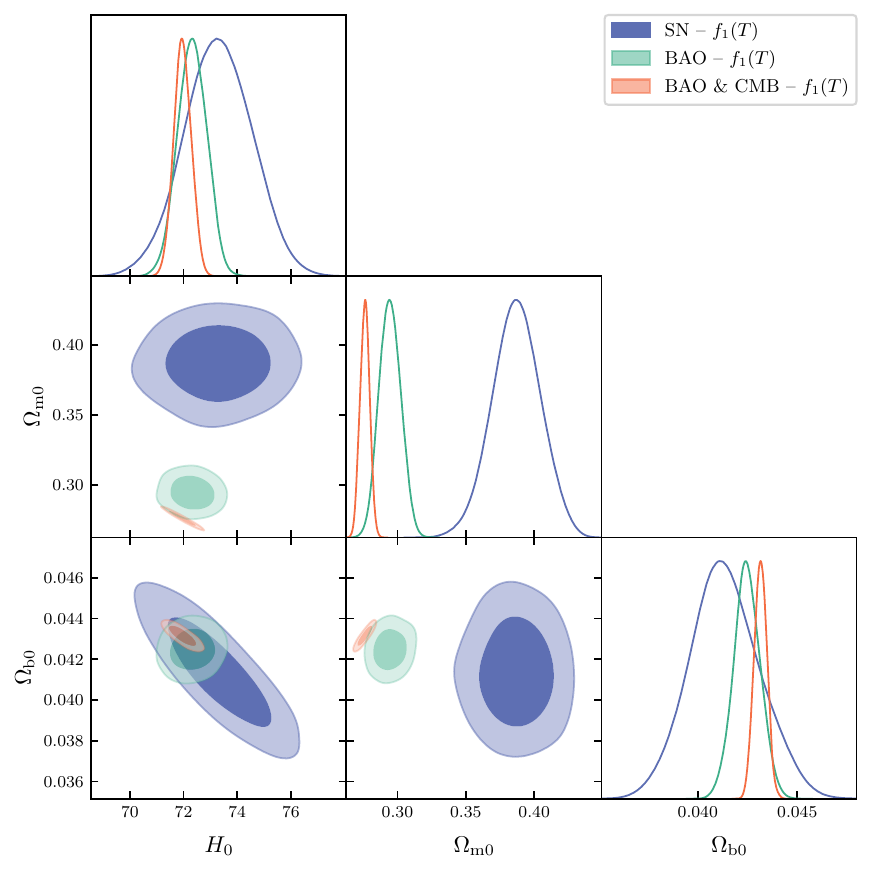} \\
        \includegraphics[width=0.49\linewidth]{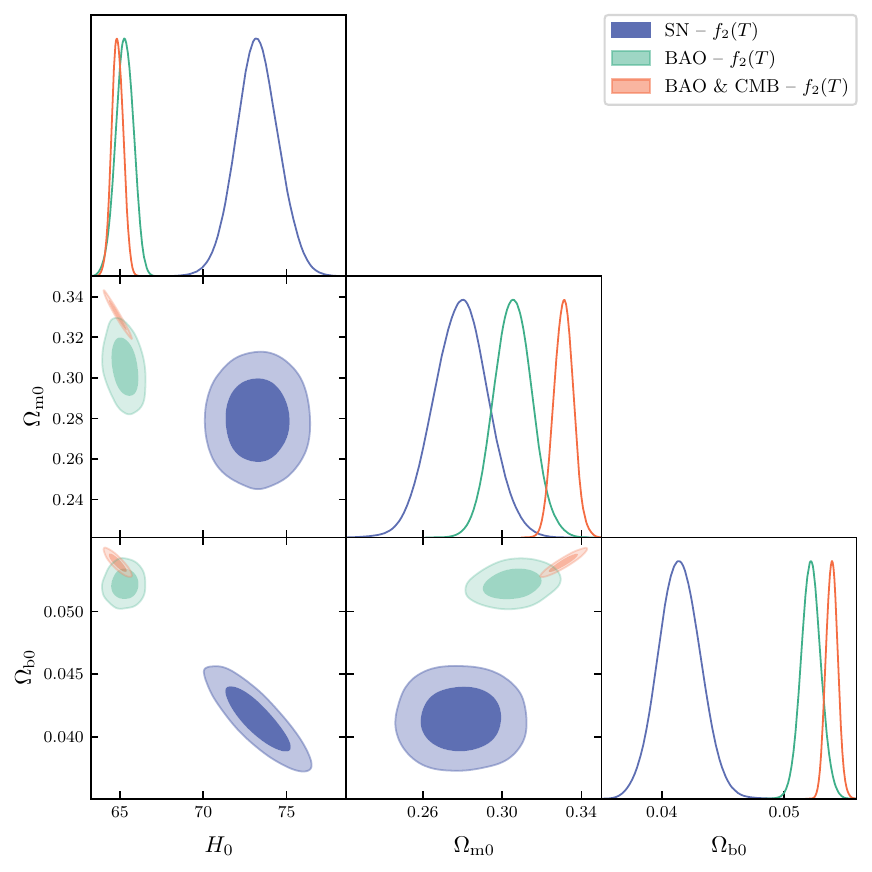} &
        \includegraphics[width=0.49\linewidth]{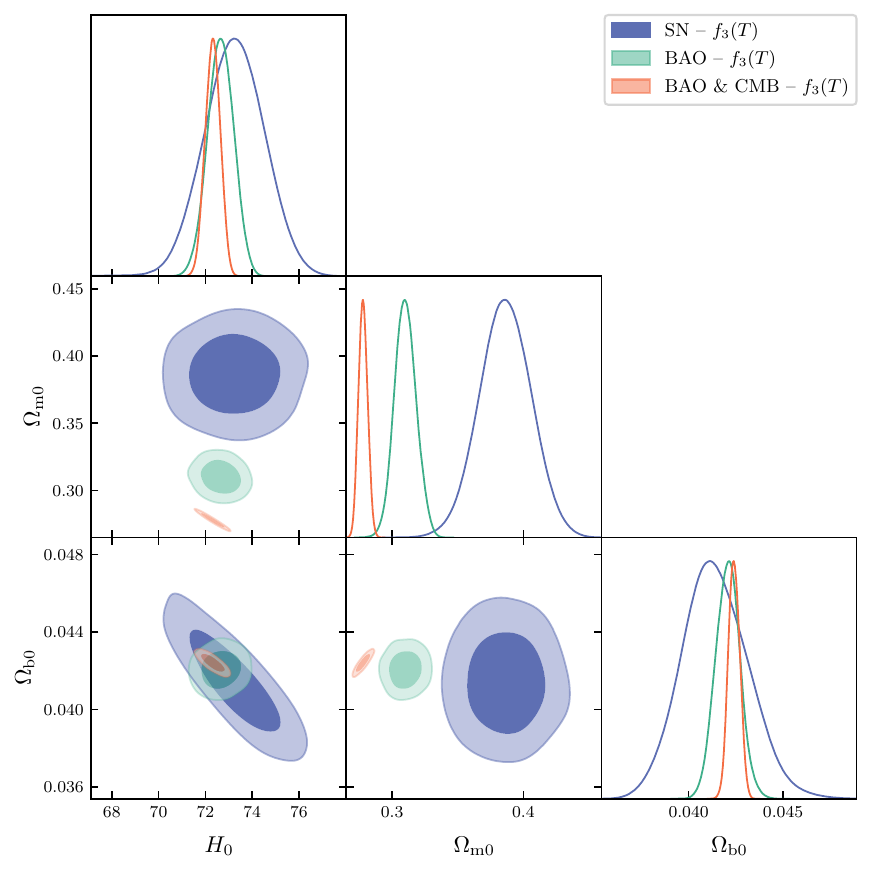}
    \end{tabular}
    \caption{\justifying{{\it{Comparison of the two-dimensional posterior distributions in the $(H_0,\Omega_{\mathrm{m}0})$, $(H_0,\Omega_{\mathrm{b}0})$, and $(\Omega_{\mathrm{m}0},\Omega_{\mathrm{b}0})$ planes obtained from the individual SN, BAO, and BAO+CMB datasets. The contours correspond to the 68\% and 95\% confidence levels (C.L.). The top-left panel shows the results for the $\Lambda$CDM model, while the remaining panels correspond to the $f_1(T)$ (top-right), $f_2(T)$ (bottom-left), and $f_3(T)$ (bottom-right) models. The relative displacement and overlap of the contours reveal the presence of internal tensions among the datasets for each model.
    }}}}

    \label{fig:tensions_fT}
\end{figure*}

\begin{figure}[t]
    \centering
    \includegraphics[width=0.49\textwidth]{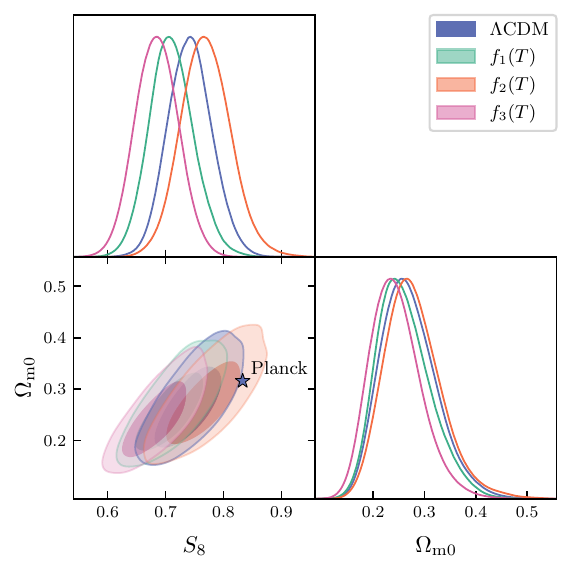}
    \caption{\justifying{{\it{
    Two-dimensional posterior distributions in the $(\Omega_{\mathrm{m}0},S_8)$ plane obtained from the RSD dataset.
    The contours correspond to the 68\% and 95\% confidence levels (C.L.) for the $\Lambda$CDM and $f(T)$ gravity models.
    For reference, the Planck 2018 best-fit constraint \cite{Planck:2018vyg} is also shown.
    }}}}
    \label{fig:rsd_S8_Om}
\end{figure}

\begin{figure*}[t]  
    \centering
    \includegraphics[width=0.95\textwidth]{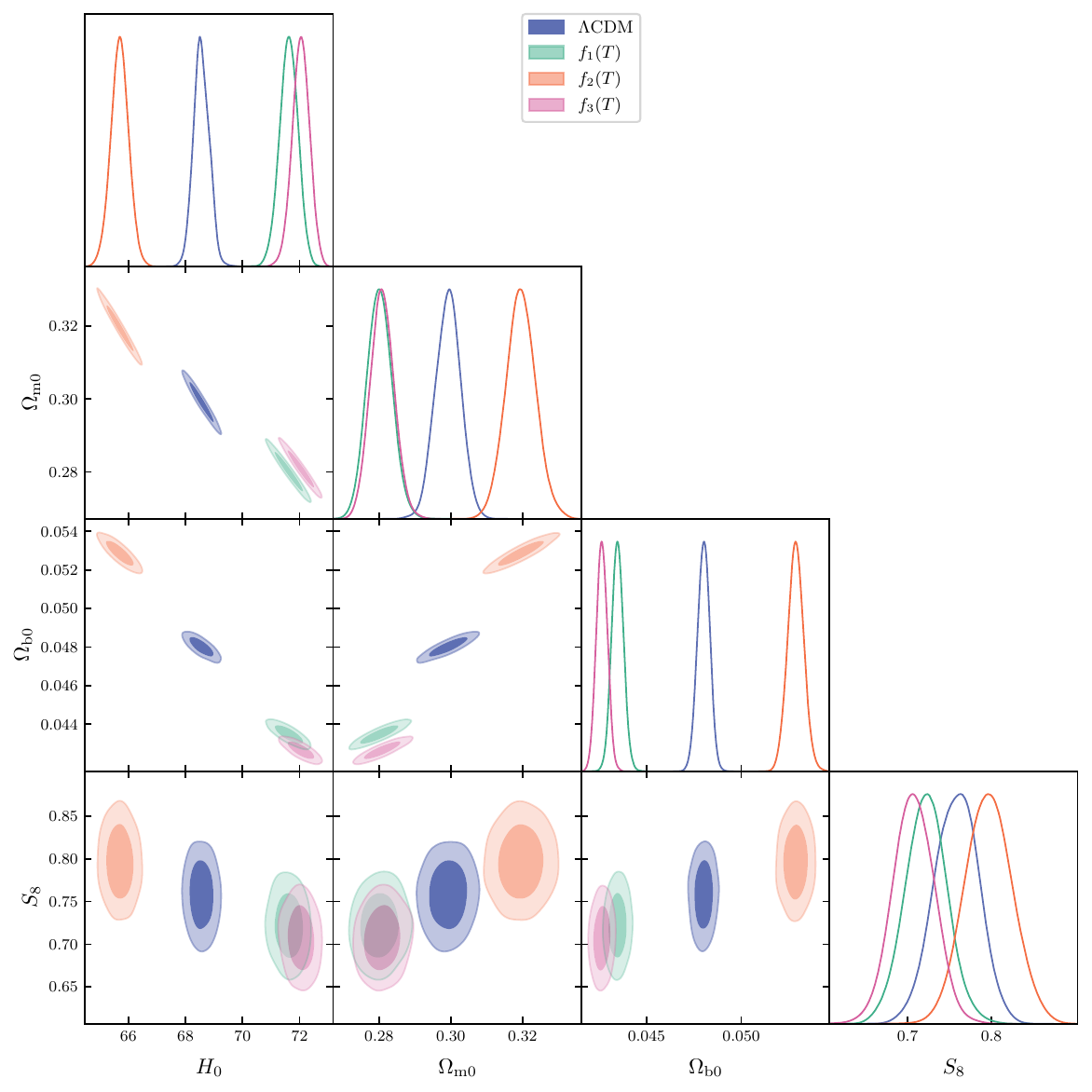}  
    \caption{\justifying{{\it{
    Two-dimensional posterior distributions for the $\Lambda$CDM and $f(T)$ gravity models obtained from the full dataset combination (SN + BAO + CMB + RSD).
    The contours correspond to the 68\% and 95\% confidence levels (C.L.).
    The figure displays all one- and two-dimensional marginalised constraints among the parameters
    $H_0$, $\Omega_{\mathrm{m}0}$, $\Omega_{\mathrm{b}0}$, and $S_8$.
    Differences in the location of the contours illustrate how the various $f(T)$ parametrisations modify the joint parameter constraints relative to the $\Lambda$CDM scenario.
    }}}}
    \label{fig:full_constraints_fT}
\end{figure*}

In this subsection we present the constraints on the cosmological parameters of
the three $f(T)$ models introduced in Sec.~\ref{fTmodel}, and we compare them
with the reference $\Lambda$CDM scenario. We report the mean values and
$1\sigma$ uncertainties obtained from the MCMC analysis for each dataset
separately and for the full combination, together with the statistical
performance quantified through $\Delta\mathrm{AIC}_{\mathrm{C}}$ (see
Sec.~\ref{infocrit}). The numerical results are summarised in
Table~\ref{tab:fT_results}. In addition, the relative consistency of different
datasets within each model is illustrated in Fig.~\ref{fig:tensions_fT}. When RSD data are analysed alone, the
constraints in the $(\Omega_{\mathrm{m}0},S_8)$ plane are displayed in
Fig.~\ref{fig:rsd_S8_Om}, while
the full joint constraints from the combined dataset are shown in
Fig.~\ref{fig:full_constraints_fT}.

\paragraph{Supernovae (SN).}
For the SN-only analysis we employ Pantheon+ as an unanchored dataset and we
calibrate the absolute magnitude through the Gaussian $H_0$ prior described in
Sec.~\ref{sndata}. Due to the imposed local calibration, the inferred values of
the Hubble constant are nearly identical across all models, with
$H_0\simeq73~\mathrm{km\,s^{-1}\,Mpc^{-1}}$. The main model-dependent impact
appears in the inferred matter density: while $\Lambda$CDM prefers
$\Omega_{\mathrm{m}0}=0.3321\pm0.0186$, the $f_2(T)$ model favours a lower value,
$\Omega_{\mathrm{m}0}=0.2792\pm0.0136$, whereas $f_1(T)$ and $f_3(T)$ push toward
higher matter density, $\Omega_{\mathrm{m}0}\approx0.386$. Since supernovae
constrain only the redshift dependence of the luminosity distance, they do not
provide direct information on the baryon density. Therefore, in this case the
constraints on $\Omega_{\mathrm{b}0}$ are entirely driven by the Gaussian BBN
prior adopted on $\Omega_{\mathrm{b}0}h^2$, which explains why all models yield
very similar values for this parameter in Table~\ref{tab:fT_results}. From a
statistical point of view, all $f(T)$ models remain compatible with
$\Lambda$CDM for SN data alone, with no statistically significant preference for
any specific parametrisation.

\paragraph{Baryon acoustic oscillations (BAO).}
Using DESI DR2 BAO measurements (Sec.~\ref{baodata}), we obtain strong
constraints on the matter density and on the BAO distance scale. In practice,
BAO primarily constrain $\Omega_{\mathrm{m}0}$ and the degenerate combination
$H_0 r_{\rm d}$, rather than $H_0$ and the sound horizon $r_{\rm d}$
separately. The BBN-motivated Gaussian prior on $\Omega_{\mathrm{b}0}h^2$ fixes
the early-Universe physics entering $r_{\rm d}$ and therefore breaks this
degeneracy, allowing us to infer $H_0$ and $\Omega_{\mathrm{b}0}$ individually.
In $\Lambda$CDM we find $H_0=68.645\pm0.505$ and
$\Omega_{\mathrm{m}0}=0.29747\pm0.00861$. The $f(T)$ models exhibit a markedly
different behaviour in the inferred Hubble constant: $f_1(T)$ and $f_3(T)$ shift
$H_0$ upwards to $H_0\simeq72$--$73$, while $f_2(T)$ shifts it downwards to
$H_0=65.278\pm0.529$ (Table~\ref{tab:fT_results}). The matter density remains
similar among all models, with $\Omega_{\mathrm{m}0}\simeq0.30$. Since BAO do not
directly constrain the baryon density, the imposed BBN prior implies an
anticorrelation between $\Omega_{\mathrm{b}0}$ and $H_0$, with larger $H_0$
favouring smaller $\Omega_{\mathrm{b}0}$. From a statistical perspective, BAO
data alone do not show a significant preference for a specific $f(T)$
parametrisation: $f_2(T)$ remains compatible with $\Lambda$CDM
($\Delta\mathrm{AIC}_{\mathrm{C}}=-1.2$), while $f_1(T)$ and $f_3(T)$ are
disfavoured, with $\Delta\mathrm{AIC}_{\mathrm{C}}\simeq 9.8$ and $10.1$,
respectively (Table~\ref{tab:jeffreys}).

\paragraph{Baryon acoustic oscillations and cosmic microwave background (BAO+CMB).}
Combining BAO measurements with CMB distance priors (Sec.~\ref{cmbdata}), we
obtain very tight constraints on the physical matter and baryon densities. In
contrast to the SN and BAO-only cases, CMB data directly constrain both
$\Omega_{\mathrm{m}0}h^2$ and $\Omega_{\mathrm{b}0}h^2$, with the latter being
significantly tighter than the BBN prior adopted in this work. This reduces the
parameter uncertainties and induces an anticorrelation between $H_0$ and both
$\Omega_{\mathrm{b}0}$ and $\Omega_{\mathrm{m}0}$, such that models with larger
$H_0$ favour smaller matter and baryon density parameters. In the $\Lambda$CDM
scenario we find $H_0=68.401\pm0.292$ and
$\Omega_{\mathrm{m}0}=0.30111\pm0.00374$, while the $f(T)$ models separate into
two behaviours: $f_1(T)$ and $f_3(T)$ favour larger values of the Hubble
constant, $H_0\simeq72$, with smaller matter densities, whereas $f_2(T)$ yields a
lower value, $H_0=64.850\pm0.348$, with a larger $\Omega_{\mathrm{m}0}$
(Table~\ref{tab:fT_results}). From a statistical perspective, BAO+CMB data
strongly disfavour all three $f(T)$ models relative to $\Lambda$CDM, with
$\Delta\mathrm{AIC}_{\mathrm{C}}=11.6$, $13.6$, and $27.3$ for $f_1(T)$, $f_2(T)$,
and $f_3(T)$, respectively (Table~\ref{tab:fT_results}).

\paragraph{Redshift-space distortions (RSD).}
Redshift-space distortion (RSD) measurements constrain the growth of cosmic structures through
$f\sigma_8(z)$ (Sec.~\ref{rsddata}), and are therefore primarily sensitive to the matter density and clustering amplitude, rather than to the background expansion history. As a consequence, RSD data alone do not break the degeneracy imposed by the Gaussian BBN prior on $\Omega_{\mathrm{b}0}h^2$, and do not provide direct constraints on $H_0$ or $\Omega_{\mathrm{b}0}$, which are therefore not reported in Table~\ref{tab:fT_results}. We find $\Omega_{\mathrm{m}0}=0.2695\pm0.0540$ and $S_8=0.7423\pm0.0378$ in $\Lambda$CDM, while the $f(T)$ models yield broadly consistent matter densities but systematic shifts in the clustering amplitude: $f_1(T)$ and $f_3(T)$ favour lower values, $S_8=0.7104\pm0.0385$ and $0.6848\pm0.0371$, respectively, whereas $f_2(T)$ yields a slightly larger value, $S_8=0.7695\pm0.0430$. These differences reflect the distinct growth histories induced by the effective gravitational coupling in each model. From a statistical point of view, all $f(T)$ models remain compatible with $\Lambda$CDM for RSD data alone, with $\Delta\mathrm{AIC}_{\mathrm{C}}<0.2$, indicating that RSD measurements by themselves cannot distinguish between the models.

When comparing the constraints obtained from the different datasets, it is
important to emphasise that the presence of cosmological tensions is not a
peculiarity of $f(T)$ gravity, since the $\Lambda$CDM model itself exhibits
well-known inconsistencies, most notably in the determination of the Hubble
constant. What changes in the $f(T)$ scenarios is the way in which these
tensions are redistributed among cosmological parameters. In particular, for
those models that shift the inferred value of $H_0$ towards local measurements,
the residual mismatch does not disappear but is instead transferred to the
matter sector. This behaviour is clearly illustrated in
Fig.~\ref{fig:tensions_fT}, where models that alleviate the $H_0$ tension exhibit
a noticeable discrepancy in the inferred values of $\Omega_{\mathrm{m}0}$
between early- and late-time probes, even when the corresponding $H_0$
constraints become more compatible.

\begin{figure*}[t]
    \centering
    \begin{tabular}{cc}
        \includegraphics[width=0.5\textwidth]{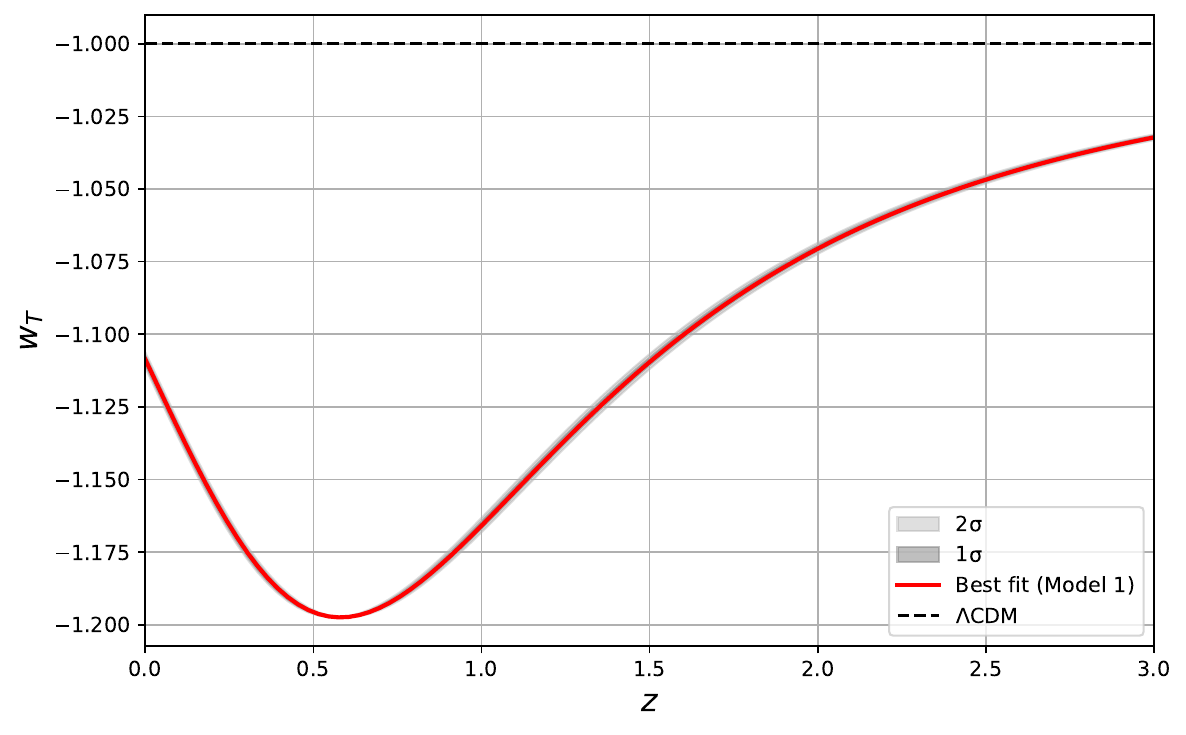} &
        \includegraphics[width=0.5\textwidth]{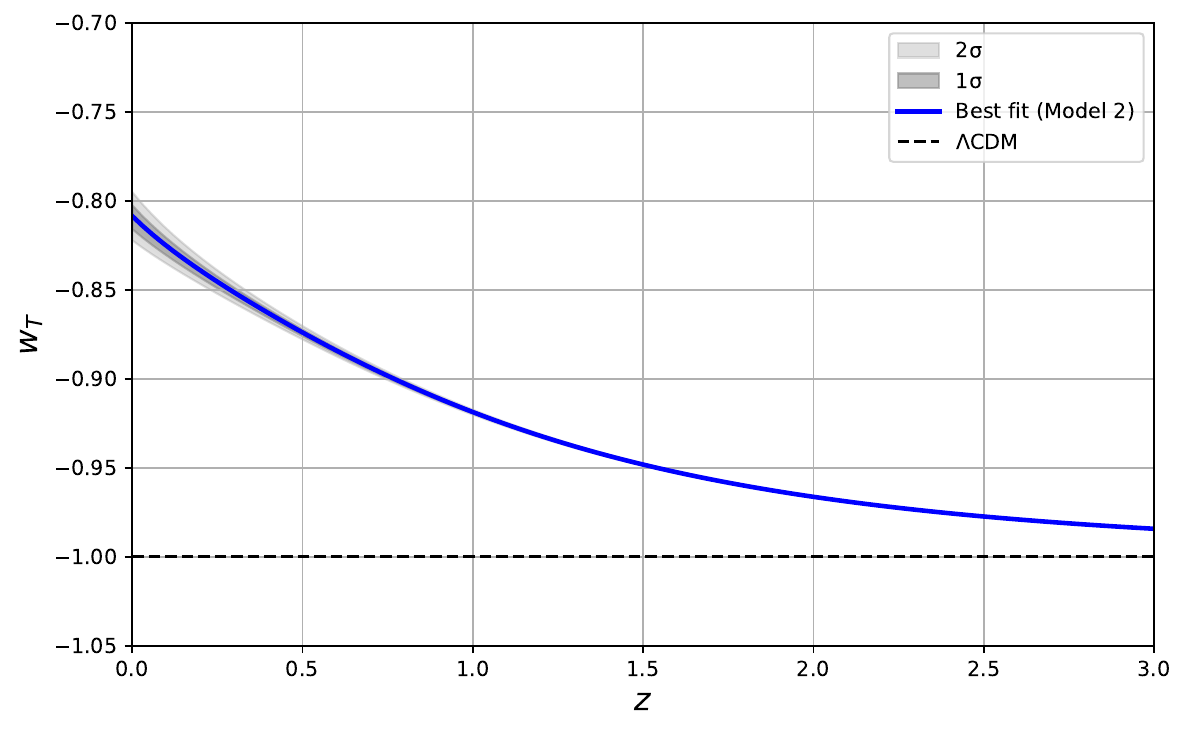} \\
        \multicolumn{2}{c}{\includegraphics[width=0.6\textwidth]{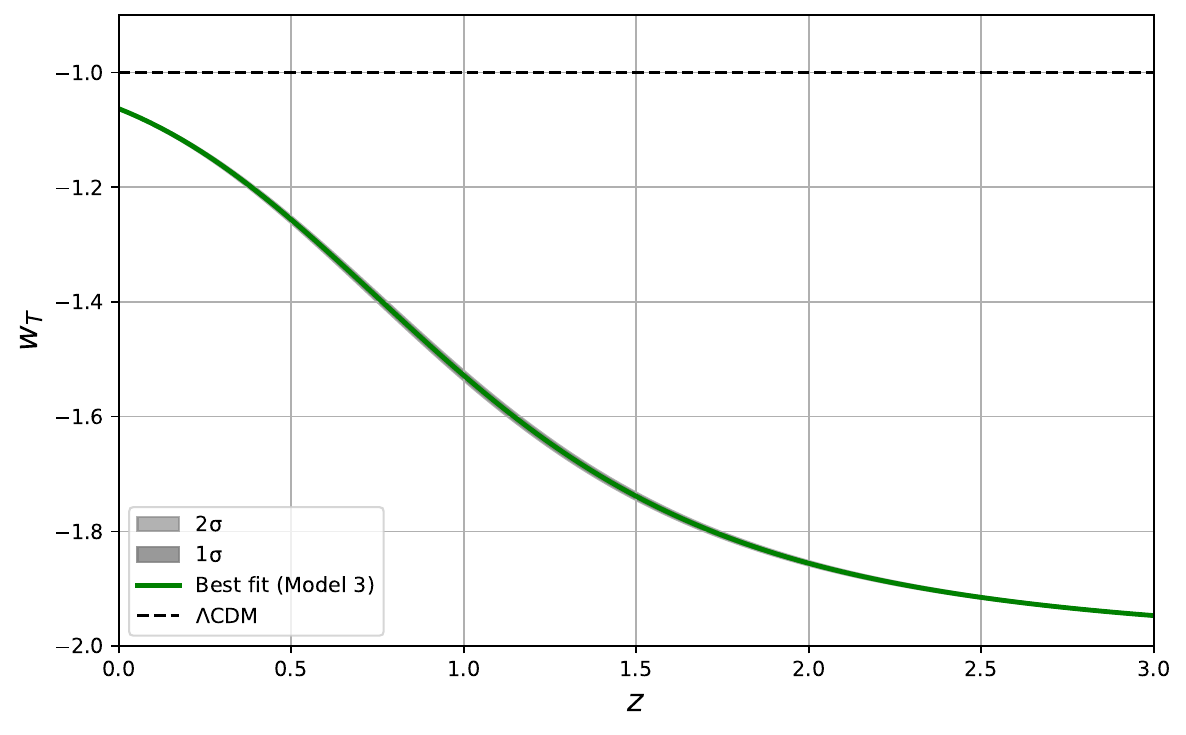}} \\
    \end{tabular}
    \caption{\justifying{{\it{
    Reconstruction of the effective dark energy equation of state parameter $w_T(z)$ for the three $f(T)$ models using the full dataset combination SN+BAO+CMB+RSD. Solid curves correspond to the best-fit reconstruction, while shaded regions represent the 68\% (dark gray) and 95\% (light gray) CL obtained from the MCMC analysis. 
    The three panels correspond to: Model~1 (top-left), Model~2 (top-right), and Model~3 (bottom). 
    Compared to Fig.~\ref{fig:theory_wde}, which displayed representative $w_T$ evolutions for each $\lambda$ parametrisation, the results presented here incorporate the full statistical confidence intervals, demonstrating that the constraints on 
    $w_T$ are consistently tight across all three models.}}}}
    \label{fig:wde_reconstr}
\end{figure*}

Focusing on the implications for the $H_0$ tension, two qualitatively distinct
classes of behaviour emerge within the $f(T)$ framework. The $f_1(T)$ and
$f_3(T)$ models consistently shift the BAO- and CMB-inferred values of the
Hubble constant towards larger values compared to $\Lambda$CDM, bringing them
closer to local distance-ladder determinations. In contrast, the $f_2(T)$ model
pushes $H_0$ to smaller values, thereby exacerbating the discrepancy. This
dichotomy can be traced back to the effective torsional equation of state:
models exhibiting a phantom-like regime ($w_T<-1$) over the relevant redshift
range enhance the late-time expansion rate and therefore favour larger inferred
values of $H_0$, whereas models with a quintessence-like behaviour ($w_T>-1$)
have the opposite effect (see Figs.~\ref{fig:theory_wde} and~\ref{fig:wde_reconstr}). As a result, in the models that partially alleviate
the $H_0$ tension, the remaining inconsistency manifests primarily in the
matter density inferred from different datasets.

Before addressing the implications for the $S_8$ sector, it is worth noting
that recent weak-lensing analyses, such as the latest KiDS results \cite{Wright:2025xka}, report no
statistically significant tension with $\Lambda$CDM. Nevertheless, this issue
must still be carefully examined in the context of modified gravity models,
since departures from General Relativity typically alter the effective
gravitational coupling and may reintroduce a mismatch between early- and
late-time probes even when none is present in the standard scenario. This point
is particularly relevant in theories such as $f(T)$ gravity, where the growth
of structures is modified through an effective gravitational coupling,
$G_{\rm eff}=G/f_T$.

Within this framework, the $f_2(T)$ model, characterised by $G_{\rm eff}<G$,
yields a larger value of $S_8$ when inferred from RSD data, while at the same
time one would expect a smaller value of $S_8$ inferred from CMB observations,
potentially improving the consistency between early- and late-time estimates.
In contrast, the $f_1(T)$ and $f_3(T)$ models correspond to $G_{\rm eff}>G$,
leading to an enhancement of structure growth and therefore to a situation in
which discrepancies in the $S_8$ sector can arise even if none is present in
$\Lambda$CDM. This behaviour is reflected in the RSD constraints shown in
Fig.~\ref{fig:rsd_S8_Om}, where the different models populate distinct regions
in the $(\Omega_{\mathrm{m}0},S_8)$ plane (see also Fig.~\ref{fig:theory_wde}).

Taken together, these results reveal a clear complementarity between the
$H_0$ and $S_8$ sectors within the minimal $f(T)$ framework. Models that are
more successful in shifting $H_0$ towards locally measured values tend to
transfer the residual tension to the matter density and to the growth of
structures, while the model that exhibits a potentially more favourable
behaviour in the $S_8$ sector aggravates the $H_0$ discrepancy. This trade-off
illustrates the difficulty of simultaneously addressing both tensions within
simple $f(T)$ extensions of the standard cosmological model, and suggests that
more general formulations or additional degrees of freedom may be required to
achieve a fully consistent resolution.

\paragraph{Full dataset combination.}
Finally, we consider the full dataset combination SN+BAO+CMB+RSD, for which the
joint constraints are shown in Fig.~\ref{fig:full_constraints_fT} and the
numerical results are summarised in Table~\ref{tab:fT_results}. In this case,
the behaviours identified in the separate analyses are combined and clearly
reflected in the multidimensional parameter space. In $\Lambda$CDM we obtain
$H_0=68.559\pm0.278$, $\Omega_{\mathrm{m}0}=0.29923\pm0.00354$,
$\Omega_{\mathrm{b}0}=0.048014\pm0.000326$, and $S_8=0.7582\pm0.0263$, while the
$f(T)$ models follow the trends discussed above. The $f_1(T)$ and $f_3(T)$ models
shift the inferred Hubble constant to larger values,
$H_0\simeq71.6$ and $72.0$, while favouring lower $\Omega_{\mathrm{m}0}$,
$\Omega_{\mathrm{b}0}$, and $S_8$, whereas the $f_2(T)$ model yields a lower
$H_0\simeq65.7$ and higher values of $\Omega_{\mathrm{m}0}$,
$\Omega_{\mathrm{b}0}$, and $S_8$. The interplay between background and growth
constraints tightens the posterior distributions and reveals the
redistribution of tensions discussed above, with partial alleviation of the
$H_0$ tension accompanied by increased discrepancies in the matter and
clustering sectors. These combined effects are clearly visible in
Fig.~\ref{fig:full_constraints_fT}. From a statistical perspective, the full dataset combination disfavors all three $f(T)$ parametrisations relative to $\Lambda$CDM, yielding $\Delta\mathrm{AIC}_{\mathrm{C}}=39.0$, $44.6$, and $46.3$ for $f_1(T)$, $f_2(T)$, and $f_3(T)$, respectively (Table~\ref{tab:fT_results}).

 We close this section by presenting in Fig.~\ref{fig:wde_reconstr}, the full statistical reconstruction of $w_T(z)$ for the three f(T) models, obtained from the MCMC chains using our last combined datasets. These plots extend the illustrative best-fit curves by explicitly including the $1\sigma$ (dark gray band) and $2\sigma$ (light gray band) confidence regions. The narrow width of the reconstructed bands over $0 \le z \le 3$, indicates that the effective dark energy equation of state parameter is strongly constrained by the data, permitting only small deviations from the best-fit curves.

 In particular, the reconstruction for Model~2 (blue curve) reinforces its characterization as a quintessence-like model ($w_T>-1$), as the entire $2\sigma$ confidence interval remains above the $\Lambda$CDM limit ($w = -1$) throughout the late-time evolution. Conversely, Models 1 and 3 (red and green curves, respectively) are statistically confirmed to reside in the phantom-like regime ($w_T<-1$). This consistency between the best-fit behaviours and the full statistical error bands reveals the limited freedom available to these $f(T)$ modifications and solidifies the distinct cosmological signatures identified for each model.

\begin{figure}[t]
    \centering
    \includegraphics[width=0.49\textwidth]{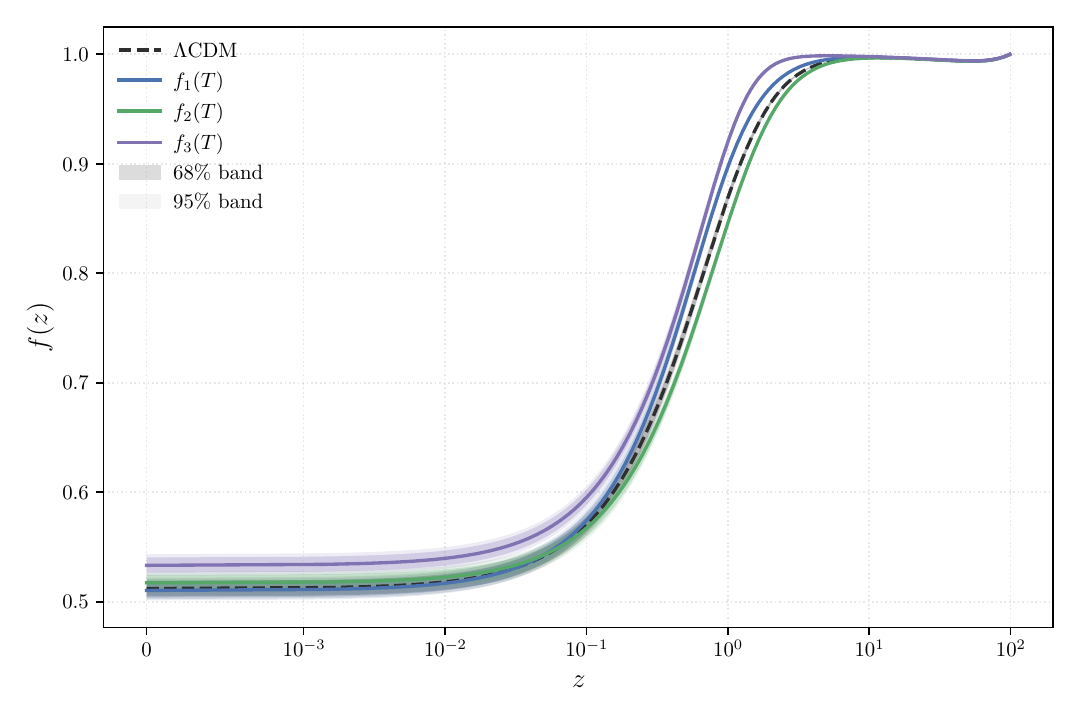}
    \caption{\justifying{{\it{
    Reconstruction of the linear growth rate 
    $f(z)=d\ln\delta_{\rm m}/d\ln a$ obtained from the full dataset
    combination SN+BAO+CMB+RSD.
    Solid curves correspond to the best-fit evolution for the three
    $f(T)$ models, while the shaded regions represent the 68\% (darker)
    and 95\% (lighter) confidence levels derived from the MCMC analysis.
    The dashed black curve shows the $\Lambda$CDM prediction for comparison.
    Although the effective torsional equation of state $w_T(z)$ and the
    effective gravitational coupling $G_{\rm eff}(z)$ can differ from their
    standard values at late times, the parameter values preferred by the
    data lead to growth histories that remain compatible with the
    $\Lambda$CDM evolution over the redshift range probed by current observations.
    }}}}
    \label{fig:growth_reconstr}
\end{figure}

We also reconstruct the linear growth rate
$f(z)$ defined in \eqref{growthrate} for the three $f(T)$ models.
We numerically solve the linear growth equation derived in Section~\ref{fTtheory}
in the subhorizon (quasi-static) regime.
The integration is started at $a_{\rm ini}=0.01$, deep in the matter-dominated era,
and the growing-mode solution for $\delta_{\rm m}$ is propagated up to $a=1$.
From this evolution we obtain $f(z)$ and construct the best-fit curve together
with the 68\% and 95\% C.L.\ bands, shown in Fig.~\ref{fig:growth_reconstr}.
This reconstruction corresponds to the same growth evolution entering the
RSD observable $f\sigma_8(z)$.

The parameters entering this reconstruction are those preferred by the full
MCMC inference. For these statistically favoured regions of parameter space,
the resulting evolution of $f(z)$ remains broadly consistent with the
$\Lambda$CDM prediction over the redshift interval probed by current data.
This occurs even though the effective torsional equation of state $w_T(z)$
can depart from $-1$ at late times and the effective gravitational coupling
$G_{\rm eff}=G/f_T$ can differ from $G$.
Hence, the allowed modifications in both the background and perturbation
sectors combine in such a way that the overall linear growth history remains
compatible with current large-scale-structure observations.

\section{Conclusions}
\label{conclusions}

In this work we investigated whether late-time modifications of gravity in the
teleparallel framework can contribute to the resolution of the current
$H_0$ tension. Focusing on $f(T)$ cosmology, we considered three representative
parametrisations, Eqs.~(\ref{f1}), (\ref{f2}), and (\ref{f3}), which reduce to the teleparallel equivalent
of General Relativity at early times and deviate from it only at late epochs. The first two models have been previously discussed in the literature, while the third model, $f_3(T)$, represents a novel $f(T)$ parametrisation inspired by a similar functional form previously studied in the context of $f(Q)$ gravity~\cite{Boiza:2025xpn}.
For each model we derived the background cosmological equations and adopted an
effective torsional-fluid description, allowing us to characterise the
torsional sector through an evolving equation-of-state parameter $w_T(z)$.
Whenever redshift-space distortion information is included, we additionally
accounted for the modified linear growth of matter perturbations, which in
$f(T)$ gravity is governed by an effective gravitational coupling
$G_{\rm eff}=G/f_T$ and a vanishing gravitational slip at linear order within
the quasi-static subhorizon approximation.

We confronted the three models with a set of late- and early-time cosmological
probes, including Pantheon+ Type~Ia supernovae treated as unanchored (calibrated
through a local $H_0$ prior), DESI DR2 BAO measurements, Planck-based compressed
CMB distance priors, and a compilation of $f\sigma_8$ measurements. The
resulting constraints are summarised in Table~\ref{tab:fT_results}, and the
dataset consistency and combined posteriors are shown in
Figs.~\ref{fig:tensions_fT}-\ref{fig:full_constraints_fT}.

Our analysis reveals two qualitatively distinct behaviours within the
considered $f(T)$ scenarios. The $f_1(T)$ and $f_3(T)$ models systematically
shift the BAO- and BAO+CMB-inferred value of the Hubble constant to larger
values than in $\Lambda$CDM, bringing the inferred $H_0$ closer to local
distance-ladder measurements. In contrast, the $f_2(T)$ model shifts $H_0$
downwards and therefore worsens the discrepancy. This dichotomy is naturally
understood in terms of the effective torsional dynamics: parametrisations that
realise a phantom-like behaviour of the torsional fluid over the relevant
redshift range tend to enhance the late-time expansion rate and hence favour a
larger inferred $H_0$, whereas a quintessence-like effective regime produces the
opposite effect. However, even in the cases where the $H_0$ discrepancy is
partially alleviated, the improvement is not achieved for free: the residual
inconsistency is largely transferred to other sectors, most notably to the
matter density inferred from early- and late-time probes, as illustrated in
Fig.~\ref{fig:tensions_fT}.

Concerning the growth of structures, the three $f(T)$ models modify the
effective gravitational coupling, $G_{\rm eff}=G/f_T$, leading to systematic
shifts in the inferred clustering amplitude. The $f_1(T)$ and $f_3(T)$ models,
which partially alleviate the $H_0$ tension, predict lower values of $S_8$ from
RSD data, while the $f_2(T)$ model favours a higher $S_8$. This behaviour is
directly related to $G_{\rm eff}$: $f_1(T)$ and $f_3(T)$ correspond to
$G_{\rm eff}>G$, enhancing structure growth and thus requiring a smaller
late-time normalisation of fluctuations, but implying larger CMB-inferred values
of $S_8$ and hence an increased early--late discrepancy. Conversely, $f_2(T)$
features $G_{\rm eff}<G$, suppressing growth and potentially reducing the
$S_8$ tension while worsening the $H_0$ one. This complementarity illustrates
the difficulty of simultaneously addressing both tensions within minimal
$f(T)$ scenarios.

Finally, we assessed the global statistical performance of the models relative
to $\Lambda$CDM using the corrected Akaike Information Criterion. Although some
parametrisations are able to shift the inferred value of $H_0$ towards local
measurements, the combined dataset yields positive values of
$\Delta\mathrm{AIC}_{\mathrm{C}}$ for the full SN+BAO+CMB+RSD combination
(Table~\ref{tab:fT_results}), indicating that the minimal $f(T)$ extensions
considered here are not statistically favoured over the reference scenario.
Nevertheless, these models provide a concrete demonstration that late-time
torsional modifications of gravity can non-trivially impact both the background
expansion and the growth of cosmic structures, and can partially redistribute
the current cosmological tensions among different sectors.

In addition to cosmological constraints, viable modified gravity scenarios must also satisfy local and strong-field tests, such as those arising from Solar-System experiments and observations of binary systems. In the present work, however, we have examined the considered $f(T)$ models exclusively at cosmological scales. The distinct behaviours identified for the effective gravitational coupling, $G_{\rm eff}=G/f_T$, provide a first qualitative indication of how these parametrisations may depart from General Relativity. A dedicated analysis of their implications for parametrised post-Newtonian constraints or binary dynamics lies beyond the scope of the present study and will be explored in future work.

In summary, our results reveal teleparallel gravity as a promising and
well-controlled framework for exploring extensions of the standard cosmological
model in light of present observational tensions. While the simple one-parameter
$f(T)$ parametrisations studied in this work do not simultaneously resolve the
$H_0$ and $S_8$ tensions, they clearly illustrate the rich phenomenology
offered by torsional modifications of gravity. This motivates the investigation
of more general teleparallel scenarios, including extended torsional
Lagrangians, non-minimal couplings, or additional degrees of freedom, as well
as a refined treatment of observational systematics. Such directions offer
promising avenues for future research.

\acknowledgments
MB-L is supported by the Basque Foundation of Science Ikerbasque.
CGB acknowledges financial support from the FPI fellowship PRE2021-100340 
of the Spanish Ministry of Science, Innovation and Universities. 
This work was supported by the Spanish grant PID2023-149016NB-I00 (funded by MCIN/AEI/10.13039/501100011033 and by “ERDF A way of making Europe"). This research is also supported by 
the Basque government Grant No. IT1628-22 (Spain).  
The authors acknowledge the contribution of the COST Action CA21136 “Addressing 
observational tensions in cosmology
with systematics and fundamental physics (CosmoVerse)”.

\bibliography{bibliografia}

\end{document}